# Advancement of metal oxide nanomaterials on agri-food fronts

Georges Dubourg[a], Zoran Pavlović[a], Branimir Bajac[a], Manil Kukkar[a], Nina Finčur[b], Zorica Novaković[a], Marko Radović[a]

[a] University of Novi Sad, Center for Sensor Technologies, Biosense Institute, Dr Zorana Đinđića 1,21000 Novi Sad, Serbia

[b] University of Novi Sad Faculty of Sciences, Department of Chemistry, Biochemistry and Environmental Protection, Trg Dositeja Obradovića 3, 21000 Novi Sad, Serbia

**Highlights:**

- Metal oxide nanomaterials are applied in food control monitoring, food packaging and agricultural production.

- Metal oxide nanomaterials are applied in agricultural environments for wastewater treatment and soil remediation.

- The risks associated with deploying $MO_x$ nanomaterials in the agri-food sector are not fully addressed.

**Graphical abstract**

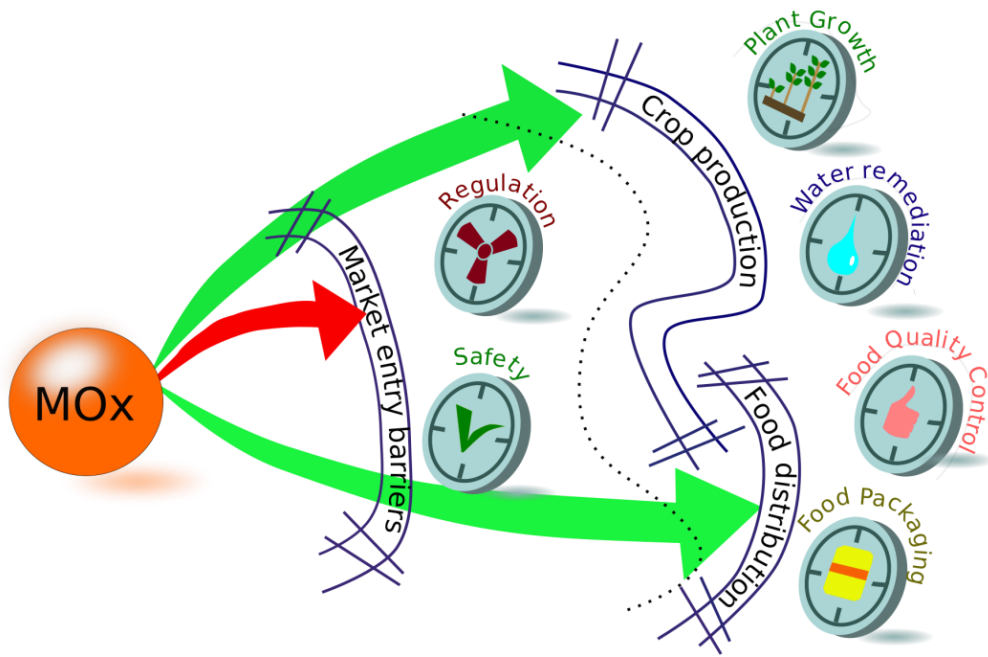

**Abstract**

The application of metal oxide nanomaterials ($MO_x$ NMs) in the agrifood industry offers innovative solutions that can facilitate a paradigm shift in a sector that is currently facing challenges in meeting the growing requirements for food production, while safeguarding the environment from the impacts of current agriculture practices. This review comprehensively illustrates recent advancements and applications of $MO_x$ for sustainable practices in the food and agricultural industries and environmental preservation. Relevant published data point out that $MO_x$ NMs can be tailored for specific properties, enabling advanced design concepts with improved features for various applications in the agrifood industry. Applications include nano-agrochemical formulation, control of food quality through nanosensors, and smart food packaging. Furthermore, recent research suggests $MO_x$'s vital role in addressing environmental challenges by removing toxic elements from contaminated soil and water. This mitigates the environmental effects of widespread agrichemical use and creates a more favorable environment for plant growth. The review also discusses potential barriers, particularly regarding $MO_x$ toxicity and risk evaluation. Fundamental concerns about possible adverse effects on human health and the environment must be addressed to establish an appropriate regulatory framework for nano metal oxide-based food and agricultural products.

## 1. Introduction

Nanotechnology is a key enabling technology that involves manipulating NMs at the atomic or molecular level. Their small size provides them with unique advantages compared to larger-scale counterparts, including remarkable physical, chemical, and optical properties with a large surface-to-volume ratio. Nanoparticles (NPs), which are just a few nanometers in size, can be found in a wide range of consumer products like deodorants, toothpastes, and paints. They have the potential to transform various industries such as medicine, aerospace, and energy production. Today, NMs are being employed in the "agri-food sector" as a means to tackle the urgent challenges of increasing crop productivity to meet the ever-growing demands of the global population, while simultaneously reducing the environmental impact caused by current agricultural practices (Dhankhar and Kumar, 2023). In the last decade, numerous reviews have outlined the beneficial effects of NMs on the growth, yield, and protection of agricultural crops (Jafir et al., 2023, Verma et al., 2018; Shang et al., 2019; Singh et al., 2022), as well as their potential applications for food quality control and preservation (Torres-Giner et al., 2020; Shafiq et al., 2020; Ghosh et al., 2019; Jagtiani, 2022). Among NM, metal-oxide ($MO_x$) nanostructures offer many advantageous features.

They can take on various structural geometries, exhibit metallic, semiconductor, or insulator characteristics, and possess unique physicochemical and photocatalytic properties (Song et al., 2015). These $MO_x$-nanostructures, whose properties can be adjusted to fulfill specific objectives, offer viable solutions to address the challenges currently faced in various aspects of the agrifood industry such as crop cultivation, distribution, and food quality control. For instance, the use of $MO_x$-based nanosensors, including electrochemical and biosensors, has become a prominent research method for sensing applications (George et al., 2018; Liu and Liu, 2019; Krishna et al., 2022). These advanced nanosensors can detect contaminants like food toxins, monitor flavor production, and identify pathogens, allowing for remote monitoring of food quality and safety with heightened effectiveness Galstyan et al., 2018, Ungureanu et al., 2022).

Besides being used for monitoring and ensuring the quality of food, $MO_x$-NMs possess the capacity to enhance plant growth and protect plants from diseases. The unique characteristics of $MO_x$, including their small dimensions, large surface area, enhanced solubility, and antimicrobial properties, make them highly valuable in the development of fertilizers and pesticides (Hyder et al., 2023). Additionally, $MO_x$ materials can positively contribute to the process of soil remediation by efficiently purifying polluted soils from hazardous elements (Ali et al., 2023). This results in providing a fertile ground for agricultural practices. Furthermore, their excellent photocatalytic characteristics and photochemical stability can support the provision of clean and abundant freshwater, which is essential for agriculture (Singh, et al., 2023 ). This, in turn, can help reduce groundwater pollution caused by the use of other agricultural chemicals like pesticides or fertilizers (Pattnaik et al., 2023).

However, despite the significant potential of $MO_x$-based NMs for sustainable and viable agriculture with a reduced environmental impact, they remain controversial due to their potential hazards to ecosystems and human health. There is a lack of understanding of the effects of $MO_x$-NMs on the environment and human health, which limits the development of regulations and legislation needed for the smooth transition of NM-based technologies into marketable products for the agri-food industry.

This review article provides industry stakeholders and the research community with an overview of the latest advancements achieved in $MO_x$-based NMs on the agri-food front. The article focuses on two distinct aspects: 1. the main front line, which highlights the positive implications of $MO_x$-NMs on several important sectors of the agri-food chain, such as crop productivity, food supply and environment, and 2. the back line, which defines the actual market barriers. In this section, we discuss the lack of knowledge regarding the impact of NMs on human and environmental health and how this affects the regulatory framework.

## 2. $MO_x$ NMs for food control and preservation

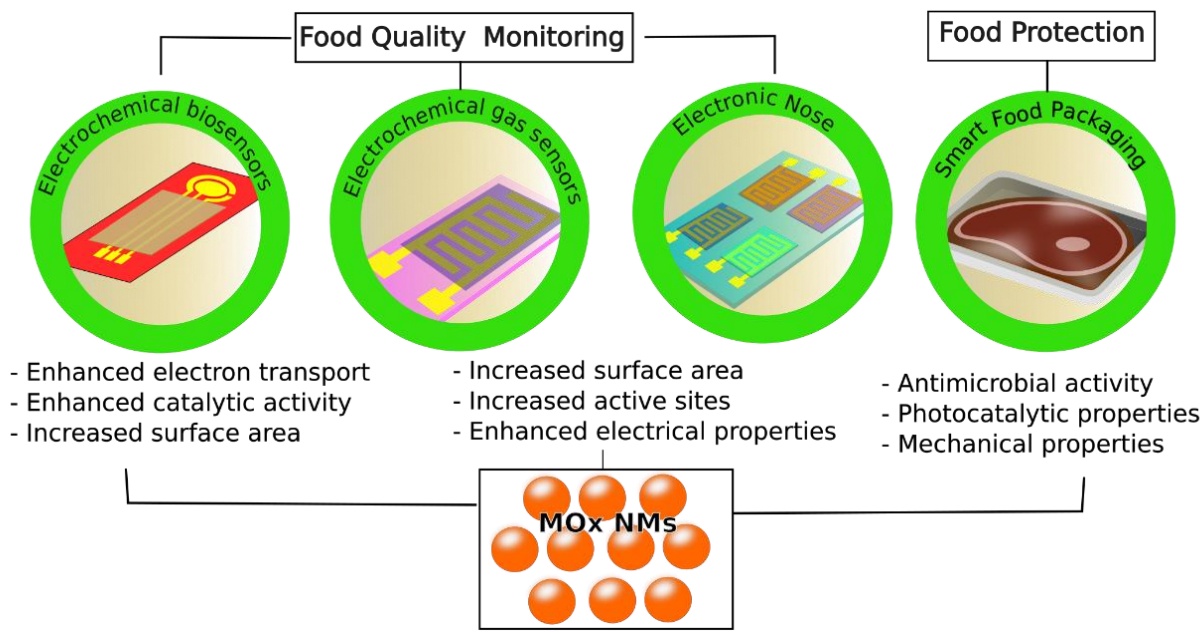

**Figure 1.** Schematic illustration of the role of $MO_x$ in the quality food control and monitoring process

The quality and safety of food must be continuously monitored and controlled from processing and storage until the products reach the consumer's hands. This is done in order to protect human health against the risks of food poisoning and to potentially reduce food waste. The process involved in ensuring food safety is complex and encompasses the monitoring of various factors, including freshness, authenticity, toxin levels, presence of pathogenic bacteria, and adulterants, at every stage of the food supply chain. Additionally, efforts are made to develop intelligent food packaging that can protect food and prolong its shelf life. The utilization of $MO_x$ NMs can play a significant role in monitoring and preserving food quality throughout the food supply chain.

In this section, we explore various types of $MO_x$ NMs that are utilized in the development and design of novel electrochemical nanosensors, electronic noses (E-nose), and nanobiosensors for the purpose of monitoring food quality. Additionally, the discussion will highlight the application of $MO_x$ NMs in food packaging, specifically their antimicrobial and photocatalytic properties that aid in safeguarding food products from potential contamination as illustrated in Figure 1.

## 2.1. Electrochemical Biosensors

Biosensors represent a wide group of sensors that have been adopted by the food safety market due to their high reliability, speed precision and user-friendly nature. The increase in the demand for sensors in various diversified fields has boosted scientists and engineers to prioritize the development of mobile devices for on-the-go analysis, which are rapid, intuitive and economical as compared to traditional laborious laboratory methods.

Key areas of the food safety value chain where biosensors based on NMs have found application include the detection of food borne pathogens, cell metabolites, heavy metal ions, toxins (pesticides, herbicides,), soil fertilizers (potassium (K), phosphorus (P), nitrogen (N)), drugs, flavors, and sweeteners. The most desired ones should be scaled down to a single chip (lab on a chip), for multiple detection purposes (Dkhar et al., 2023). A bioreceptor and a transducer are the two major components of an electrochemical biosensor. The bioreceptor is responsible for the recognition of the target analyte whereas the transducer transforms this recognition event into a quantifiable signal (Naresh and Lee, 2021). An analyte represents any micro or macro biomolecule, ion, heavy metals, toxins, bacteria, viruses etc. Depending on the analyte of interest, bioreceptor (biological recognition element) must be precisely chosen. Bioreceptors most commonly used are the enzymes, antibodies and aptamers as nucleic acid fragments, (Kukkar et.al., 2018, Huang et al., 2021, Podunavac et.al., 2023, Novakovic et.al., 2024). A typical construction of a biosensor with different types of receptors is shown in Figure 2.

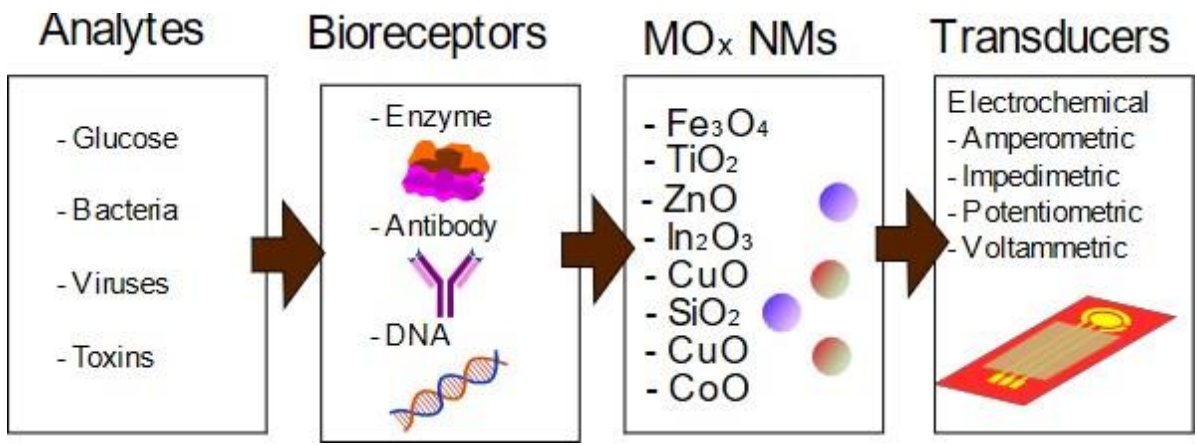

**Figure 2.** Schematic illustration of general concept of an electrochemical biosensor based on $MO_x$ NMs

The type of transducer determines the design of the sensor, and they can be broadly categorized as piezoelectric, optical, mechanical, acoustic, magnetic, and electrochemical, based on the detection principle involved (Hariri et al., 2023). Electrochemical biosensors are the most versatile and highly developed among chemical and biochemical sensors. The electrical signal is generated through the conversion of biochemical processes or electrochemical redox reactions. Enhancements in sensitivity and specificity of electrochemical sensors are achieved through the use of various materials such as noble metal and $MO_x$ NPs, carbon-based NMs, transition metal dichalcogenides, nitrides and carbides, ion-selective membranes, and field-effect transistors as signal amplifiers. Functionalized NM-based electrocatalysts can enhance the sensitivity of electrochemical sensors due to their unique electrocatalytic properties and by expanding the active electrochemical surface area where the reaction occurs (Li et al., 2013; Aftab et al., 2023). The electrochemical signals that are given to the electrode and measured as feedback signals, determine the type of sensor, which includes voltammetric, potentiometric, amperometric, conductometric, impedimetric, and coulometric sensors (Saputra, 2023).

The advantages of using electrochemical biosensors are very good selectivity and sensitivity, reproducibility, miniature size, small sample volume required, and the possibility of simultaneous detection of more desired elements by lab on a chip design with various sensors (Novakovic et al., 2024). Other desired properties are low limit of detection (LOD), wide range of detection, rapid response, cost-effectiveness, easy to use, disposable, low power consumption, portable, reusable, easily integrable with micro-electronics, self-calibration, and self-cleaning (Grieshaber et al., 2008). Signal and data post-processing play a very important role in obtaining accurate and credible results of detection.

There are several generations of electrochemical biosensors developed yet. The first generation of biosensors detects the product of a chemically catalyzed reaction, which causes the electrochemical response at the electrode by redox reaction (e.g., hydrogen peroxide as a by-product of the enzymatic reaction). The second generation of biosensors introduces specific molecules called 'mediators' (any good redox couple like ferrocene or ferro/ferricyanide) that transfer the charge between the enzyme and the electrode while an enzymatic reaction takes place to generate the electrochemical response (Dede and Altay, 2018). In the third generation of biosensors, the electrochemical response is caused by the reaction itself; the enzyme is placed into an integrated polymer matrix, or redox couples are covalently bound to the enzyme for direct electron transfer to the electrode during the enzymatic reaction (Juska and Pemble, 2020). Non-enzymatic, so-called "fourth generation" of biosensors is based on the electrochemical signal of direct oxidation/reduction of biomolecules of interest on the NM-modified substrate, without any additional biorecognition elements and mediators (Taha et al., 2020).

The most studied enzymatic biosensors are glucose sensors, which are important detection devices in the beverage, food quality control, and fermentation manufacturing sectors. These sensors, depending on the type of enzyme used, dehydrogenase (glucose dehydrogenase (GDH)) or oxidize (glucose oxidase, graphene oxide ($GO_x$)) glucose molecules, and provide an electrical response that corresponds to the glucose concentration (Yoo et al., 2010). Enzyme immobilization is one of the crucial steps in developing highly sensitive and stable enzymatic biosensors since it affects the bioactivity of the enzymes. The creation of new NMs opens up possibilities for customizing enzyme characteristics and optimizing immobilization. In electrochemical sensorics, functional nanostructured transition $MO_x$ materials are promising in catalytic-based sensorics, due to their stability, excellent catalytic activity, cost-effectiveness, and ease of fabrication. Manganese Oxide (MnO$_2$) in combination with Multi wall Carbon nanotubes (MWCNTs) as a nanocomposite, exhibit remarkable catalytic activity and facilitates the detection of H$_2$O$_2$ compared to the individual effects of MWCNTs and MnO$_2$ nanowires. This eases the researchers to achieve a broad linear range from (5-200 µM & 0.2-1 nM) and a LOD of 2 µM (Hao et al., 2020).

A flexible polyethylene terephthalate (PET) substrate platform was proposed by Mao et al., 2021. Thick film of 50 nm of gold was deposited by Ion Sputtering and Zinc Oxide nanorods (ZnO NRs) are hydrothermally grown from ZnO seeds at temperature of 90°C for two and half hours. The surface was tailored by electrodeposited rGO sheets after a thorough analysis of the optimal number of cycles required for rGO deposition on ZnO NRs. The potential impact on electrode surface morphology and transmission of redox electrons was investigated by varying the number of deposition cycles. Twenty (20) cycles of electrodepositions give the optimal response as compared to 6 and 40 cycles for sensor fabrication with Glucose oxidase (Gox) and ion selective membrane Nafion. These glucose sensors demonstrated heightened sensitivity of 5.40 µA mM$^{-1}$ cm$^{-2}$ and a diminished LOD of 17.8 µM , with an expanded linear range spanning from 01.mM to 12 mM. Additionally, these sensors exhibited enhanced catalytic activity in the electro-oxidation of hydrogen peroxide, resulting in a lower overpotential for the reaction.

The usage of NMs, especially MO$_x$, which shows a catalytic effect for glucose oxidation, are potential candidates for on-enzymatic electrochemical sensors. Tian et al., 2018 explained how enzymatic glucose sensor performance depends on the catalytic activity and stability of enzymes (operating temperature, pH level, and humidity), which degrade over time. They found that the detection performance of nickel oxide (NiO), copper (II) oxide (CuO), and cobalt oxide (CoO) is different due to their various electrochemical activities, electronic conductivity, and catalytic behavior towards the reaction of glucose oxidation, where CuO is shown to have the highest selectivity and sensitivity for glucose detection.

Most urea biosensors need urease enzymes, which reduces their lifetime in harsh settings, however, Dutta et al. (2014) suggested an enzyme-free urea biosensor based on tin dioxide ($SnO_2$) quantum dots (QD) /rGO composite. Carbon-based material composites and QDs are integrated for the urea biosensor-sensitive transducer surface fabrication due to their exceptional ability to provide faster electron transfer. The amperometry technique showed the sensitivity of $SnO_2$ QD/rGO to urea in the range of $1.6 \times 10^{-14} - 3.9 \times 10^{-12}$ M, with a LOD of 11.7 µM.

Foodborne pathogen-contaminated food has always been a great threat to human health, and due to that, the development of immunosensors is of great importance towards healthier nutrition. Immunosensors are types of affinity biosensors using an antibody as a bioreceptor to determine and quantify specific antigens present at the surface of a targeted biological analyte (Mahato et al., 2018). Antibodies could be immobilized on the transducer by crosslinking with the carboxyl, aldehyde, amino, or sulfide groups to achieve strong covalent bonding (Yuan et al., 2023). The use of $MO_x$ NMs in electrochemical immunosensors to enhance their sensing characteristics is reflected in the amelioration of the electrical affinity and conductivity, as well as the reduction of the electrode/electrolyte interface's impedance to improve redox current. Also, $MO_x$ provides the sensor with a high surface-to-volume ratio to increase the number of biorecognition sites of the electrochemical immunosensors, thus amplifying their immunological interaction and sensitivity (Patil et al., 2023).

Barbieri et al., 2023 described a novel electrochemical immunosensor for the specific detection of aflatoxin B1 in peanuts by immobilizing polyclonal antibodies to amino-functionalized iron oxide NPs ($Fe_3O_4$-$NH_2$) to increase the robustness of the biorecognition layer. Such an approach showed a linear response with a LOD of 9.47 μg/mL and a limit of quantification of 28.72 μg/mL. Moreover, Feng et al., 2022 created a $Fe_3O_4$@graphene nanocomposite electrochemical immunosensor with electrodeposited Au NPs for the detection of *Salmonella*. This modification gave a very good electrical signal, due to graphene's mechanical stability and excellent electrical conductivity and $Fe_3O_4$'s large specific surface area, which increases the overall sensitivity of the immunosensor.

Wang et al., 2008 described precise, specific, and rapid immunosensors using electrochemical impedance spectroscopy (EIS). The immunosensor were based on titanium dioxide ($TiO_2$) nanowire bundle microelectrodes for detecting *Listeria monocytogenes*, where positively charged monoclonal antibodies were conjugated with negatively charged $TiO_2$ nanowire bundles to specifically capture *L. monocytogenes*. Both types of mentioned immunosensors showed a lower LOD of $2.4 \times 10^2$ cfu/mL for *Salmonella* and $4.7 \times 10^2$ cfu/mL for *L. monocytogenes*. A Minimalistic label free immunosensor has been recently developed, leveraging the synergistic impact of Gold Nanoparticles (AuNPs) and Titanium Oxide Nanoparticles (TiO2) on the Screen-Printed Gold Electrode. This setup, in conjunction with Nafion as a selective membrane, enables ultra-sensitive detection of β-lactoglobulin. The fusion of Tris(2,2′-bipyridyl)ruthenium(II) ([Ru(bpy)3]2+) and tri-n-propylamine (TPrA) demonstrated its effectiveness as a reliable pairing, serving dual roles as a luminophore and co-reactant, thereby facilitating the generation of the electrochemiluminescence(ECL) signal. This ECL immunosensor exhibited an impressive LOD, detecting concentrations as low as 0.01 pg/mL. It displayed two clearly defined linear ranges: 0.01–50 pg/mL and 100–750 pg/mL, each with high correlation coefficients (R2 = 0.99136 and R2 = 0.97829), respectively.

 (Hong et al., 2022a).

Aptasensors are a type of biosensor with high specificity for diverse detections, where affinity may be built for various molecular, ionic, and microbiological detections as a consequence of the ideal sequence of nucleotides. Aptamers are far more stable in terms of structural and conformational stability than antibodies and enzymes (Thiviyanathan and Gorenstein, 2012, Kukkar et.al., 2018). NMs used in the development of aptasensors may perform a variety of functions, including carrier and enhancer, which amplify the signal due to electrocatalytic activity, high surface-to-volume ratio, and favorable electronic properties, then as catalysts (nanoenzyme-like), which exhibit enzyme-like activities, reporters, separators (e.g.magnetic NPs), and quenchers (Saleh and Hassan, 2023). Various quencher nanomaterials, such as AuNPs, may be employed in aptasensors based on electrogenerated chemiluminescence quenching. Also, the usage of aptamers as biorecognition elements may provide the required selectivity and specificity to improve the detection of heavy metals, because of their strong affinity and selectivity towards a variety of ions and biomolecules (Zhan et al., 2016).

The interaction between Ti\O groups and phosphate groups on deoxyribonucleic acid (DNA) aptamers embedded using electrostatic interactions over the rGO-TiO$_2$ nanocomposite for detection of *Salmonella enterica serovar Typhimurium* was explored by Muniandy et al., 2019. Using the differential pulse voltammetry (DPV) technique, the variations in electrical conductivity of bacteria against their specific aptamer were assessed, where the bacteria act as a resistance that interrupts electron transmission, resulting in a decrease in the DPV signal in proportion to the reduction in bacterial cell concentrations. This aptasensor demonstrated strong selectivity for *Salmonella* bacteria, great sensitivity, a wide detection range ($10^8$ to $10^1$ cfu mL$^{-1}$), and a low LOD of $10^1$ cfu mL$^{-1}$.

Prabhaka et al., 2016 developed a chitosan/fluorine doped tin oxide (FTO) aptaelectrode doped with iron oxide that was cross-linked to the biotinylated DNA aptamer sequence unique to malathion, using streptavidin as the bridge. The assay time with the aptaelectrode is 15 minutes with a LOD of 1 pg/ml even without sacrificing selectivity. The applicability of the proposed aptasensor was effectively assessed in real samples of soil and lettuce leaves, with a recovery of 80-92% recovery of malathion. According to Liu et al., 2019a, a highly sensitive and specific electrochemical aptamer-based sensor has been created for the detection of microcystin-LR by combining molybdenum disulfide ($MoS_2$)-coated $TiO_2$ nanobeads with Au NPs to form the ternary composite to which microcystin-LR (MC-LR) aptamers were bound by Au-sulfur links. The obtained ternary composite had better catalytic and electron transport abilities, and $TiO_2$ nanobeads coated with $MoS_2$ nanosheets provided a sizable surface area for AuNPs and biomolecules immobilization. Following that, for binding to the immobilized aptamer, MC-LR and biotin-cDNA with a complementary sequence to the aptamer created by MC-LR engaged in a competition. The current signal is catalyzed by avidin-HRP (Horseradish peroxidase) dropped as MC-LR increased and a LOD of 2 pM was established.

Prabhakar et al., 2016 used streptavidin as a linking molecule to immobilize a malathion-specific biotinylated DNA aptamer sequence on a chitosan/FTO/iron oxide electrode for malathion pesticide detection. This chitosan-iron oxide nanocomposite with immobilized aptamer biosensors showed a LOD of around 1 pg/mL, and about 80–92% recovery of malathion from soil samples and lettuce leaves.

Regarding soil quality control, reliable NPK sensors are urgently needed on many critical control points, like biochemical and microbiological contamination of plants, plant fruits, and finished food products, soil quality control (NPK fertilizers), soil and plant pesticide and herbicide contamination, inorganic contamination of soil (e.g., by heavy metals) and environmental contamination of water and air (Yin et al., 2021). In combination with other new NMs and detection elements, transition $MO_x$ offer exciting potential for such applications due to their specific electrochemical properties. Excess fertilizer has accumulated in soil over the last few decades, resulting in degradation of soil quality and toxic consequences for animals and people. As a result, a nitrite detection method based on carbon electrodes doped with CoO NPs using cyclic voltammetry (CV) method exhibited extraordinary electro-catalytic sensitivity for nitrite ion detection with a LOD of 0.3 µm and a large linear range (Puspalak et al., 2022). According to Bonyani et al., 2015, Ag/iron oxide nanocomposite-based electrochemical sensor for nitrate detection is proposed, where superior performance of two-step chemically produced silver at iron oxide, as an electrocatalyst for nitrate reduction in neutral pH was achieved. The Ag/iron oxide screen-printed electrode showed an increased cathodic current in the reaction with nitrates compared to the single-phase-modified electrode, achieving good sensitivity and a 30 mM LOD. It is possible that the uniform distribution of Ag particles on the surface of the amorphous iron oxide is responsible for the significant improvement in electrocatalytic efficiency.

Field-effect transistor (FET)-based potassium sensor was fabricated by Ahn et al., 2018a by the immobilization of valinomycin on iron (III) oxide ($Fe_2O_3$) NPs-modified ZnO nanorods. $Fe_2O_3$ NPs addition showed highly enhanced sensitivity as compared to FET devices with ZnO NRs, due to high surface area of ZnO NRs modified by $Fe_2O_3$ NPs, where valinomycin immobilization increases. $Fe_2O_3$ NPs provide ZnO surface stability in the alkaline and acidic solutions, which is useful for the detection of potassium.

Pesticide and fertilizer usage, as well as mining and burning of fossil fuels, are possible sources of heavy metal ions at trace levels in different sources, including natural waters, soil, plants, food, beverages, etc. (Ariño et al., 2017). Most of them, when present in higher amounts, are known to be toxic and carcinogenic, and due to that, reliable onsite detection is in high demand. The World Health Organization has classified organophosphorus pesticides (OPPs) as extremely dangerous neurotoxic chemical compounds which are often utilized for the crop's protection from pests (Bhattu et al., 2021). Thus, due to their intensive use and high toxicity, a high-performance electrochemical sensor for the monitoring of pesticide contamination is urgently needed. Electrochemical detection of toxins by using various $MO_x$ NPs and $MO_x$-metal nanocomposites as the electrochemical sensing platform gives promising results in sensitivity, rapidity, high surface area, and high selectivity.

A nonenzymatic electrochemical sensor for carbofuran (CBF) and carbaryl (CBR) compound detection based on CoO on rGO was devised by Wang et al., 2014, in which they showed that it is possible to detect both carbamate pesticides simultaneously. The sensor displayed a low LOD of 4.2 µg/L for CBF and 7.5 µg/L for CBR, and a linear relationship is obtained over a wide concentration window of 0.2–70 µM (R =0.9996) for CBF and 0.5–200 µM (R = 0.9995) for CBR. Gao et al., 2019 created a robust and accurate electrochemical sensor for methyl parathion pesticide residues in Chinese cabbage samples using the Au/ zirconium dioxide ($ZrO_2$)/graphene/ glassy carbon electrode (GCE) material. Due to their strong affinity for phosphoric groups, $ZrO_2$ NPs exhibit remarkable selectivity in detecting organophosphorus pesticides (OPs), and the presence of phosphate groups and $ZrO_2$ enhances the adsorption of methyl parathion on the electrode surface. Zirconia or graphene-modified electrodes (Au/$ZrO_2$-graphene/GCE) showed a better electro-catalytic response towards MP oxidation compared to the Au NPs where linear current with the concentrations of MP was in the range of 1-2400 ng /mL, with the LOD of 1 ng/mL.

Many research groups have reported productive results about the electrochemical detection of heavy metal ions with various $MO_x$ catalytic NMs, doped with other metal ions, or even as a composite NM. One of the most common forms of iron oxide used to detect heavy metals is $Fe_3O_4$, but there are only a few reports using iron oxide without any other additional ions, because iron oxide NPs may become non-conductive due to aggregation. Most studies functionalize iron oxide NPs or combine them with other materials, which produce spinel ferrites that can enhance more sensitive detection of heavy metals.

Studies done by Zhou et al., 2015 on manganese ferrite ($MnFe_2O_4$) to detect different heavy metals showed a good linear response of arsenite (As (III)) concentrations between 10 and 100 ppb, with a LOD was 1.95 ppb and a sensitivity of 0.295 μA/ppb. In a subsequent investigation, glassy carbon electrodes were modified with $MnFe_2O_4$ and graphene oxide for improved analytical performance in the detection of metal ions such as copper ions (Cu (II)), lead cation (Pb (II)), mercury cation (Hg (II)), and cadmium cation (Cd (II)), demonstrating LODs of sub-mM concentration in river water analysis (Zhou et al., 2016). Wei et al., 2012 reported linear ranges of 20 to 140 nM for cadmium (Cd) and 1 to 30 nM for lead, with very low sensitivities and limits of detection in the 10-12 M and 10-11 M ranges using magnesium oxide (MgO) nanoflowers modified glassy carbon electrodes. For metal ion detection purposes, many researchers used certain ion-selective membrane cocktails to cover the working electrode to selectively detect the ions of interest.

QDs are nano-sized particles with a size of 1 to 10 nm with many exceptional properties owing to their structural differences, due to quantum mechanical phenomena, such as large specific surface areas, excellent electrical properties, abundant active sites, easy functionalization, and good aqueous dispersibility. A range of two-dimensional (2D) QD NMs has been created using materials such as graphene, nitrides, transition $MO_x$ and dichalcogenides, Mxenes, and black phosphorus (Zhang et al., 2022a).  Much research has been conducted on using various types of QDs with adjustable surface modifications to detect ions, biomolecules, and living cells. With their high affinity for biomolecules, QDs have been widely applied in electrochemical biosensing as electrode modifiers, electron transfer accelerators, and carriers of sensitive elements, for enhancing the selectivity and sensitivity of detecting the desired analyte.

Field Effect Transistor (FET) electrochemical sensors hold significant promise for a wide range of detection applications. In FET biosensors, bioreceptors are immobilized on a semiconductor channel or sensing material connecting the source and drain electrodes. The interface of this material plays a critical role in the transduction process of FET sensors. A bias voltage is applied to the semiconductor material, allowing for modulation of its electronic properties, including electrical conductivity, by a third electrode (gate). The captured analytes induce changes in the material's conductance through mechanisms such as electrostatic gating and/or Schottky barrier modulation, resulting in a measurable signal that enables determination of analyte concentration (Sedki et.al.,2021).

FET configuration may be back-gated wafer-based with source and drain electrodes, while as a back gate, the bulk wafer could directly function (Pham et al., 2019). Many 2D and three-dimensional (3D) NMs may be grown on a wafer of dielectric material (e.g., silicon dioxide ($SiO_2$)) deposited on a conducting substrate (e.g., Si). Another possible FET structure liquid-ion gate-based, with the gate function of the ionic liquid. At the liquid-channel interface a double-layer is created, at which the electric change happens. Back-gate capacitance may be many times smaller in comparison to an ionic-liquid gate capacitance via various substrates (Lieb et al., 2019).

$MO_x$, as the most varied types of solids, may be good candidates for the library of 2D materials used for transistor fabrication with high electron mobilities higher than 10 cm$^2$ V$^{-1}$ s$^{-1}$, and a large band gap energy range of 2.3–4.9 eV. All those properties guarantee high sensitivity and low signal-to-noise ratio in biosensing applications. $MO_x$ have been applied a lot in bio/chemical detection as electrochemical transducers. Some of the most used are tungsten trioxide ($WO_3$), molybdenum trioxide ($MoO_3$), tantalum trioxide ($TaO_3$), and gallium trioxide ($Ga_2O_3$) as well as $SnO_2$, ZnO, CuO, and indium trioxide ($In_2O_3$) (Sedki et al., 2020).

One of the interesting transition $MO_x$, $In_2O_3$ was used in many FET biosensors giving good sensitivity. Chen et al., 2017 developed a 2D $In_2O_3$-based non-enzymatic FET biosensor for the glucose detection with an extremely low LOD of less than 7 fM, showing the linear response over a broad dynamic range of glucose from $10^{-11}$ - $10^{-5}$ M. Boronic acid was used as a receptor of glucose molecules. The device's performance was better than that of other non-enzymatic FET sensors for glucose when it used the recognition molecule boronic acid and semiconductor channels made from carbon-based NMs. Some examples of MOx NMs-based biosensors for agrifood applications are presented in **Table 1**.

**TABLE 1.** Summary of $MO_x$ commonly used in biosensors.

| $MO_x$ composite | Analyte | Detection Technique | Linear Range | LOD | Linear Range (real sample) | LOD (real sample) | Ref. |
|---|---|---|---|---|---|---|---|
| $MnO_2$/ MWCNT | $H_2O_2$ | Amperometry | 5-200 μM & 0.2-1 nM | 2 μM | - | - | Hao et al., 2020 |
| $SnO_2$ (QD)/rGO | urea | Amperometry | $1.6\times10^{-14}$ - $3.9\times10^{-12}$ M | 11.7 μM | - | - | Dutta et al., 2014 |
| $Fe_3O_4$-$NH_2$ | aflatoxin B1 | Cyclic voltammetry/ EIS | 0.5 to 30 μg/mL | 9.47 μg/mL | - | 3.79μg/mL (peanut) | Barbieri et al., 2023 |
| $Fe_3O_4$@graphene | *Salmonella* | CV/ DPV | $2.4 \times 10^2$ - $2.4 \times 10^7$ cfu/mL | $2.4 \times 10^2$ cfu/mL | - | $1.18 \times 10^3$ cfu/mL (milk) | Feng et al., 2022 |
| $TiO_2$ | *Listeria monocytogenes, Salmonella* | EIS | $10^4$ to $10^8$ cfu/ml for *L. monocytogenes* | $2.4 \times 10^2$ cfu/mL for *Salmonella*, $4.7\times10^2$ cfu/mL for *L. monocytogenes*. | - | - | Wang et al., 2008 |
| rGO-$TiO_2$ | *Salmonella enterica* | DPV | $10^8$ to $10^1$ cfu mL-1 | $10^1$ cfu mL−1 | - | - | Muniandy et al., 2019 |
| Chitosan/ (FTO) | malathion | DPV | 0.001–10 ng/mL | 1 pg/ml | 0.01 μg/mL, − 0.001 ng/mL Lettuce/soil sample | 0.001 ng/ml | Prabhaka et al., 2016 |
| $MoS_2$/$TiO_2$nanobeads/Au NPs | microcystin-LR | DPV | 0.005–30 nM | 2 pM | - | - | Liu et al., 2019a |
| Carbon electrodes/ CoO | nitrite ion | CV | - | 0.3 μm | - | - | Puspalak et al., 2022 |
| Ag@iron oxide/SPCE | nitrate | CV/ Amperometry | - | 30 uM | - | - | Bonyani et al., 2015 |
| CoO/rGO | carbofuran (CBF) and carbaryl (CBR) | CV/DPV | 0.2-70 μM for CBF and 0.5-200 μM for CBR | 4.2 μg/L for CBF and 7.5 μg/L for CBR | 0.5–200 uM (fruit/vegetable) | 0.50 uM | Wang et al., 2014 |
| Au/ ZrO$_2$)/graphene/GCE | methyl parathion | CV/EIS/ Square wave voltammetry | - | - | 1-2400 ng /mL (cabbage) | 1 ng/mL | Gao et al., 2019 |
| Mn/Fe$_2$O$_4$ | heavy metals | Square wave anodic stripping voltammetry( SWASV)/CV | 10 - 100 ppb | 1.95 ppb for As (III) | - | 9.56 ppb (water) | Zhou et al., 2015 |
| GCE/MnFe$_2$O$_4$/GOx | Cu (II), Pb (II), Hg (II), Cd (II) | SWASV/CV/ EIS | 10 -110 ppb. | 3.37 ppb | - | sub-mM (water) | Zhou et al., 2016 |
| MgO/GCE | Pb (II), Cd (II) | CV/Stripping voltammetry | 2.1 pM for Pb and 81 pM for Cd | 3.3 to 22 nM for Pb(II) and 40 to 140 nM for Cd(II) | 1.0 to 30 nM for Pb (II) and 20-140 nM for Cd(II) (Water) | 6.07 ppb for Pb(II) | Wei et al., 2012 |
| In$_2$O$_3$ | glucose | Chronoamperometry | $10^{-11}$ - $10^{-5}$ M | less than 7 fM | - | - | Chen et al., 2017 |

The possibilities for improving electrochemical detection of foodborne pathogens, heavy metal ions as soil and water contaminants, and toxins in food production processes lie in the discovery of more sensitive, selective, stable, catalytically active, non-toxic, and biocompatible novel NMs. The approach to NM synthesis plays a crucial role in the stability of electrocatalysts and the reproducibility of detection results. By adjusting the properties of electrocatalytic NMs such as particle and pore size, morphology, effective surface area, electron transfer properties, adsorption capacity, and the detection mechanism, the LOD should be improved. Incorporating chemical functional groups onto $MO_x$ NMs can allow for the detection of target molecules like immobilized enzymes, DNA molecules, or antigen/antibody complexes. Various electrode pretreatments, enzyme/protein engineering, immobilization/conjugation strategies, and procedures involving biological recognition elements can significantly affect the LOD. The use of molecularly imprinted polymers (MIP), specially designed NMs with electrocatalytic activity similar to biorecognition elements, specific ion-selective membranes, and electronically tuned semiconductive properties of field-effect transistors could potentially lead to the detection of analytes at femto/attomolar levels, including the detection of single binding events (Lahcen et al., 2023 ).

To put certain bio-electrochemical sensor prototypes into a practical application, as a lab-on-a-chip replacement of traditional lab detecting methods, much effort still has to be done in discovering of novel 2D and 3D biofunctionalized multifunctional NMs combined with $MO_x$ and biorecognition elements, with better stability under the operational conditions, and more affordable in terms of cost-effectiveness. Nowadays, there is already extensive work being published on various functional NMs and on their potential application in electrochemical sensors, but not many sensing devices appeared in the market, even if there is a huge demand for the practical application and various sensing activities.

Emphasis should be placed on the engineering expertise required for designing and manufacturing sensors, as well as a thorough understanding of electrochemical methods and processes. It is also important to grasp the various reaction mechanisms that occur at the surface of catalytic NMs in order to develop functional and applicable devices.

## 2.2. Electrochemical gas sensors

Electrochemical gas sensors utilizing $MO_x$-NMs have garnered significant interest for detecting volatile compounds (VOCs) in food products and beverages (Wawrzyniak, 2023). Among volatile organic compounds (VOCs), amine gasses such as trimethylamine (TEA) and dimethylamine (DMA), as well as ammonia ($NH_3$) and hydrogen sulfide ($H_2S$)are crucial indicators in food quality control (Andre et al., 2022, Wang and Zeng, 2018, Wang et al., 2022a).

In contrast to electrochemical biosensors that use $MO_x$ NPs to improve sensor performance, $MO_x$ NMs in gas sensors function as the detecting component responsible for identifying the targeted gas molecules. The adsorption and desorption processes of the targeted gaseous species on the $MO_x$ structure surface, leading to changes in the material's electrical characteristics, form the basis for the sensing mechanism of $MO_x$-based electrochemical gas sensors (Wawrzyniak, 2023). Some $MO_x$ NMs like SnO, $TiO_2$, ZnO, and CuO have been utilized for the electrochemical detection of amine gases and hydrogen sulfide. Prospective sensors must exhibit outstanding gas sensitivity and selectivity along with efficient operation at ambient temperatures, in order to decrease energy consumption and off-chip instrumentation. Challenges arise when using $MO_x$ nanostructure-based sensors as they often require high operating temperatures for optimal sensitivity and selectivity. Various approaches have been explored to address this issue, such as doping, catalysts, and modifying material structure (Figure 3).

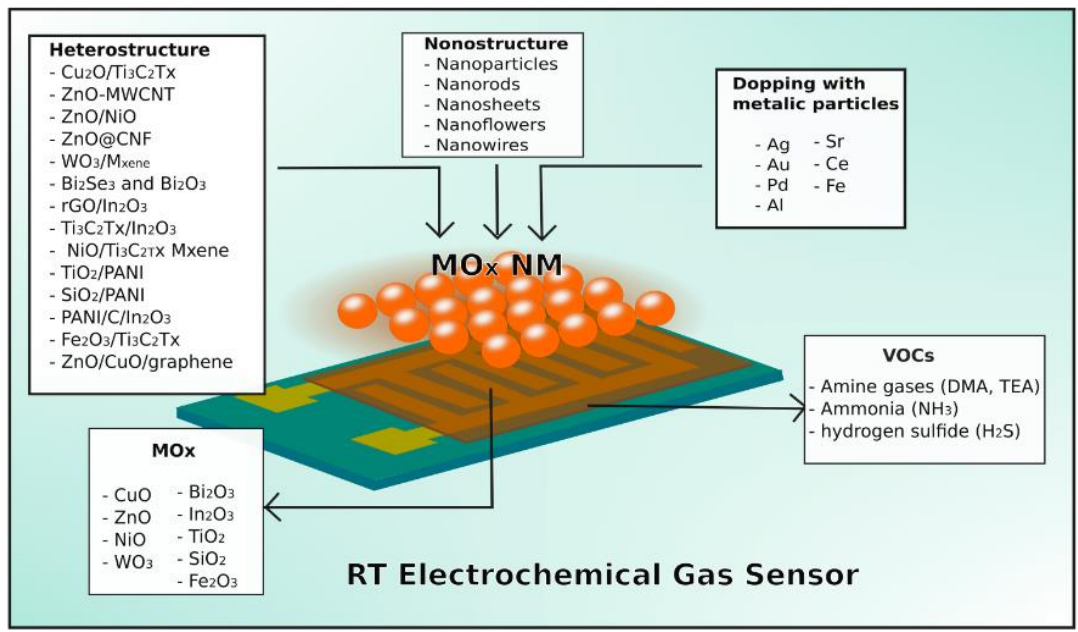

**Figure 3.** General strategies used in MOx-based electrochemical gas sensors working at RT

For instance, Xu et al., 2020 demonstrated the creation of a room temperature (RT) TMA sensor with high sensitivity to TEA by increasing oxygen vacancy concentration in a porous SnO thin layer.

The effect of exposed facet of copper (I) oxide ($Cu_2O$) NPs on the gas sensing performance toward $NH_3$ has been discussed by Zhao et al., 2023. They showed that the exposed high-index facet has superior $NH_3$ gas adsorption and surface charge activity, resulting in the enhancement of $NH_3$ sensing characteristics. Another example of sensors using $Cu_2O$ involves the formulation of $Cu_2O/Ti_3C_2T_x$ nanocomposites (Zhou et al., 2022a). The research demonstrated that this particular formulation led to a significantly greater sensitivity to triethylamine, with levels 3.5 times higher compared to the initial $Cu_2O$ nanospheres. Moreover, in order to improve sensor capabilities at RT, a variety of ZnO nanostructures were created. Srinivasan et al., 2018 conducted interesting research in which $NH_3$ sensors were created using twisted ZnO nanowires to enhance the response characteristics of the NH$_3$ sensors. The research findings demonstrated that twisted ZnO exhibits a more favorable ratio between the intensity of polar and non-polar diffraction peaks compared to other orientations of ZnO nanowires. This leads to an increased number of adsorption sites on the surface, thereby improving the gas sensor's sensitivity towards NH$_3$. Further research showed that ZnO thin films, when co-doped with Fe and aluminum (Al), demonstrate increased sensitivity to NH$_3$ in comparison to pure ZnO. This is because the doping creates a more direct path for charge carriers, thereby increasing the electrical conductivity of the sensing material (Vijayakumar et al., 2020). In a study conducted by Radhi Devi et al., 2020, a comparable method was employed utilizing zinc oxide (ZnO) doped with strontium (Sr) to develop a sensor for detecting NH$_3$ at RT. The findings revealed that the sensor incorporating Sr-doped ZnO exhibited heightened sensitivity and improved recovery properties in the presence of NH$_3$ in comparison to pure ZnO. Nanocomposite of ZnO/NiO was executed for NH$_3$ gas sensor by Jayababu et al., 2019. This methodology encompassed establishing a p-n junction between p-type NiO and n-type ZnO. The study showcased that the synergistic effect of this resultant heterojunction, in conjunction with the chemical sensitization and catalytic properties of NiO, notably augmented the sensing efficacy of the ZnO/NiO sensor at room temperature. ZnO-MWCNT was also investigated for the detection excuted on of NH$_3$ at RT (Vatandoust et al., 2021). It was shown that nanotubes fill the void between ZnO NPs and cause the ZnO surface to become rough, increasing the surface area available for gas testing.

Recently, ZnO NPs were combined with carbon nanofibers (ZnO@CNF) to produce NH$_3$ sensor working at RT (Fan et al., 2022). It was demonstrated that the working temperature of the ZnO@CNF gas sensor is significantly reduced from 325°C to 23°C compared to pure ZnO.

Additionally, ZnO nanostructure-based nanocomposites were also studied to increase the LOD for amine gasses at RT.   For instance, Bruce et al., 2022 synthesized palladium (Pd)-decorated ZnO nanoflowers on interdigitated electrodes to obtain higher response. This modification successfully increased the electron depletion layer and resistance, resulting in improved sensing capabilities. The sensor exhibited an favorable  response rate of 45% at 25 °C when exposed to 400 ppm of methylamine .

Furthermore $TiO_2$ NPs are considered as a type of $MO_x$ NM that is also highly valued for use in gas sensor applications. Various composite materials incorporating $TiO_2$ have been thoroughly studied to  develop RT gas sensors.

For instance, a ternary heterostructure gas sensor by decorating a $TiO_2/Ti_3C_2T_x$ composite with $MoO_3$ for the rapid and selective response to Triethylamine at RT (Yao et al., 2023). This approach provided both stability in humidity and sensitivity, with  a obtained response of 28%  at a TEA concentration of 5 ppm at room temperature and 80% relative humidity. In another example,  Ce doped $TiO_2$ nanocrystals were used for the detection of fish spoilage (Wu et al., 2022a). It was demonstrated that Ce doped $TiO_2$ exhibits a large specific surface area allowing the detection of $NH_3$ at RT.  Similar technique consisting in using Co-doped $TiO_2$ NPs was  investigated by Chen et al., 2022. They showed that compared with pure anatase $TiO_2$ working at 180 °C, Co-doped $TiO_2$ allows the detection of $NH_3$ at RT.  Au NP-decorated $TiO_2$ nanosheets were used as well to fabricate a high-performance $NH_3$ sensor operating at RT. They showed that Au NP-decorated $TiO_2$ nanosheets can enhance $NH_3$ sensor response at RT (Hwang et al., 2023).

Furthermore,  Zhang et al., 2022b proposed the design of a gas sensitive material by in-situ growing $(001)TiO_2$ onto a two-dimensional transition-metal carbide ($Ti_3C_2T_x$, $M_{xene}$). They introduced UV light for electron excitation, resulting in significantly increased sensitivity of the sensors for detecting $NH_3$ at room temperature for food spoilage detection. Similar methodology was studied by Yang et al., 2022. In their work, the sensing performance towards $NH_3$ of 2D polyamide (2DPI)/$In_2O_3$ composite was enhanced at RT with the irradiation of UV light.

$WO_3$, vanadium(V) oxide ($V_2O_5$) $In_2O_3$ and bismuth(III) oxide ($Bi_2O_3$) were also used for gas sensors working at RT. For instance,   passive and wireless $WO_3$/$M_{xene}$ composite sensors were proposed for the detection of TEA at RT (Li et al., 2023). It was discovered that discovered that incorporating MXene significantly enhanced the efficiency of $WO_3$ in detecting TEA at RT, resulting in a resistance increase of 277 at a concentration of 10 ppm

Liquid-phase-exfoliated layered n-$Bi_2Se_3$/p- $Bi_2O_3$ was employed in fabricating $NH_3$ sensors with excellent selectivity that work efficiently at RT, as reported by Das et al., 2023. The research findings showed that a combination of $Bi_2Se_3$ and $Bi_2O_3$ in nearly equal proportions displayed enhanced $NH_3$ sensing capabilities. This improvement was attributed to the strong interaction between the n-type $Bi_2Se_3$ and p-type $Bi_2O_3$ at ambient temperatures, resulting in an increase of 8.8 in electrical current at 180 ppm concentration of $NH_3$.

Incorporation of 3D rGO into a mesoporous $In_2O_3$ structure, as well as  $In_2O_3$ /Pd, and rGO/$In_2O_3$ composites, were introduced to enhance the sensitivity of gas sensors to TMA at RT. (Ma et al., 2019, Meng et al., 2022). Furthermore, $V_2O_5$  nanosheets were utilized in gas sensors designed for monitoring food spoilage (Van Duy et al., 2023). Due to their extremely thin thickness and porous nature, these nanosheets contribute to the sensors' effective performance in detecting $NH_3$ at RT. Furthermore, in a recent study by Zhou et al., 2022b, $Ti_3C_2T_x$/$In_2O_3$ nanocomposites were developed for RT $NH_3$ sensors, resulting in improved performance. The researchers demonstrated that the $Ti_3C_2T_x$ composite possesses a substantial surface area that facilitates gas diffusion and adsorption. Another approach for detecting $NH_3$ gas at RT involved the use of NiO/$Ti_3C_2T_x$ $M_{xene}$ nanocomposites, as proposed by Yang et al., 2023, which exhibited excellent repeatability and long-term stability. The enhanced $NH_3$ detection capability was attributed to the larger aspect ratio of the composite material, the increase of active sites introduced by NiO, and the electron interaction within the heterojunction.

An additional interesting strategy for developing electrochemical sensors that operate at RT involves combining conducting polymers such as polyaniline (PANI), which is gas sensitive at RT, with $MO_x$ like $WO_3$ (Fan et al., 2020), $SnO_2$ (Nie et al., 2018), and $TiO_2$ (Seif et al., 2019; Xiong et al., 2023) to ensure the stability of the sensor.

For the detection of DMA, Yang et al., 2021 implemented a low-temperature gas-phase diffusion technique to grow PANI on the surface of a $TiO_2$ nanorod array. The rapid deposition of the $TiO_2$/PANI nanocomposite resulted in its behavior as an n-type semiconductor. The sensor made from $TiO_2$/PANI was found to exhibit excellent selectivity, a strong response within the range of 6.36-30 ppm, and a low LOD of 0.050 ppm for DMA at RT.

The introduction of PANI was also used for the detection of $NH_3$ at RT. For instance, Aliha et al., 2023 proposed a $SnO_2$/PANI nanocomposites-based sensor for the detection of $NH_3$ at RT. They showed that the presence of polymer increases the surface area and generates a p-n junction between $n$-type $SnO_2$ and p-type PANI. A PANI/Carbon-coated hollow $In_2O_3$ Nanofiber Composite was also elaborated by Hong et al., 2022b for the sensitive detection of $NH_3$ at the RT.

$H_2S$ is another strong indicator used to assess the quality control of food and drinks (Wang et al., 2022a). Various techniques have been employed to enhance the sensitivity of $MO_x$ towards $H_2S$ at RT. One such method involves augmenting the sheet resistance of indium zinc oxide (IZO) through laser annealing, resulting in the enhancement of sensitivity of the IZO-based $H_2S$ sensor (Chang et al., 2020). Moreover, Zhang and colleagues (Zhang et al., 2019) used the synergistic effect of the CuO/MoS$_2$ binary nanostructure and charge transfer modulation within the heterojunction to increase H$_2$S sensing performance. Recently, Ding et al., 2023 constructed a heterojunction with mixed-MO$_x$ (NiO-CuO hydrangeas-like composite) to obtain large specific surface area (27.85 m2/g) and porosity (0.14 m3/g). Conducted sensing tests indicated that NiO-CuO sensors exhibit recoverable behavior and remarkable sensing performance with a LOD of 50 ppb of H$_2$S in dry air at RT. Fe$_2$O$_3$/Ti$_3$C$_2$T$_x$ derived TiO$_2$ heterostructure was considered for the low concentration detection of H$_2$S at RT (Zhang et al., 2023). In addition, Guo et al., 2022 developed H$_2$S sensor using tin(IV) selenide (SnSe$_2$)/WO$_3$ composite for monitoring egg spoilage at RT. They demonstrated that the heterojunction promotes electron migration and enhances the response to H$_2$S. Furthermore, based on the synergistic effects of the quantum size effect, surface flaws, and the large specific surface area, Wu et al., 2022b showed that TiO$_2$ QDs had a good prospective applicability in ppb-level H$_2$S detection of the volatiles related to fish spoilage. In addition, Guo et al., 2022 developed a hydrogen sulfide (H$_2$S) sensor using tin(IV) selenide (SnSe$_2$)/WO$_3$ composite for monitoring egg spoilage at RT. They demonstrated that the heterojunction promotes electron migration and enhances the response to H$_2$S. Moreover, Wu et al., 2022b showcased the potential application of titanium dioxide quantum dots (TiO$_2$ QDs) in detecting ppb-levels of H$_2$S related to the volatiles from spoiled fish, highlighting the synergistic effects of quantum size, surface flaws, and high specific surface area.

Similar approach exploited the synergistic effect of PdO and parallel nanowires assembled CuO microspheres to develop ultra-fast H$_2$S sensors at RT (Dun et al., 2022). The enhanced H$_2$S sensing properties were also improved by using the morphology control and p-n heterojunction between flower-like WO$_3$ and CuO NPs (He et al., 2019). Nanostructured copper coated on CNTs/SnO$_2$ have also been investigated for the fabrication of H$_2$S sensors and shown high sensitivity at RTs (40 ppm) ( Zhao et al., 2020).

In$_2$O$_3$ materials have been widely studied for the development of H$_2$S sensors due to In$_2$O$_3$'s unique ability to chemisorb oxygen species, facilitating its interaction with H$_2$S, a reducing gas. This interaction alters the levels of charge carriers in In$_2$O$_3$. Additionally, In$_2$O$_3$ directly interacts with H$_2$S to turn into more conductive In$_2$O$_3$.

Reducing the operational temperature of electrochemical sensors to RT enables the production of gas sensors on flexible materials, thus opening up the possibility of incorporating them into food packaging. The integration of sensor systems with communication protocols such as radio frequency identification (RFID), near field communication (NFC), and Wi-Fi in packaging could lead to new advancements in smart food packaging. This technology may allow users of smart devices to access real-time information on the quality and safety of the food. Traditional techniques, such as screen and inkjet printing, could be utilized to print wireless sensor systems on the packaging, making industrial scale production possible. For instance, Al Shboul et al., 2022 developed a printed and flexible In$_2$O$_3$ NPs-based H$_2$S sensor for Internet of things (IoT) food packaging applications. The developed sensor was successfully used for the detection of low concentrations of H$_2$S generated during the spoilage of organosulfur-rich food. Paper-based NH$_3$ sensors were also investigated using nanostructured ZnO/CuO/graphene (Jagannathan et al., 2022). In this work, ZnO/CuO nanoflowers were grown on graphene printed paper substrates. Other examples of flexible sensors for NH$_3$ detection were shown by Wen et al., 2023 and Jain et al., 2023 in which they used respectively Poly(sodium 4-styrenesulfonate) (PSS)-doped PANI/T$_{i3}$C$_2$T$_x$ composites and PANI-TiO$_2$ as NH$_3$ sensitive material. To further improve flexible sensors, more focus should be placed on the reversibility, repeatability, and resilience of such flexible systems to assure their proper operation throughout the shelf life of the food product, as well as the toxicity of $MO_x$ nanostructures in contact with the food.

Another aspect to take into consideration is that the VOCs present in food create an intricate blend of flavors formed by different compounds generated through bacterial decomposition of food. This is where gas sensors capable of detecting multiple VOCs can play a crucial role in capturing the overall aroma profile of food items. This will enable a more precise and comprehensive evaluation of the food quality. An interesting example of a multiple sensor device was proposed by Senapati and Sahu (Senapati and Sahu, 2020a and 2020b). They developed an Au metal patch electrode capacitive sensor for rapid and accurate detection of volatile gasses generated from raw fish and meat to determine their freshness status. The sensor was intricately constructed on a silicon substrate, utilizing Ag-SnO$_2$ as the sensing material on top of a SiO$_2$ layer, with the metal electrode being made of Au that allowed for detection at ppb-ppm levels of NH$_3$, TMA, DMA, and H$_2$S in fish, as well as NH$_3$, TMA, ethanol, and H$_2$S in meat.

During the process of creating various sensing devices, a promising approach involves utilizing $MO_x$-based nanosensors technology in conjunction with machine learning and AI to design advanced systems for monitoring food quality. These sophisticated systems, such as E-noses, could significantly enhance the multi-sensing capabilities in the food industry as discussed in the subsequent section.

## 2.3. E-nose application

The agrifood industry has been disrupted by the E-nose technology's rapid growth during the past decades. Significant improvements in the testing and monitoring of various food and beverage goods were made possible by the assimilation of E-nose products. The utilization of $MO_x$ materials as sensitive layers in E-nose platforms and advancements in the machine learning and artificial intelligence (AI) algorithms used for data processing are primarily responsible for the observed disruption.

The primary objective of the E-nose concept is to mimic an animal's olfactory system to detect various substances within a complex mixture of gases, scents, odors, perfumes, etc. (Figure 4). An E-nose device consists of two essential parts: a sensor system and a pattern recognition component. A typical sensing system consists of an array of sensing components and corresponding transducers.

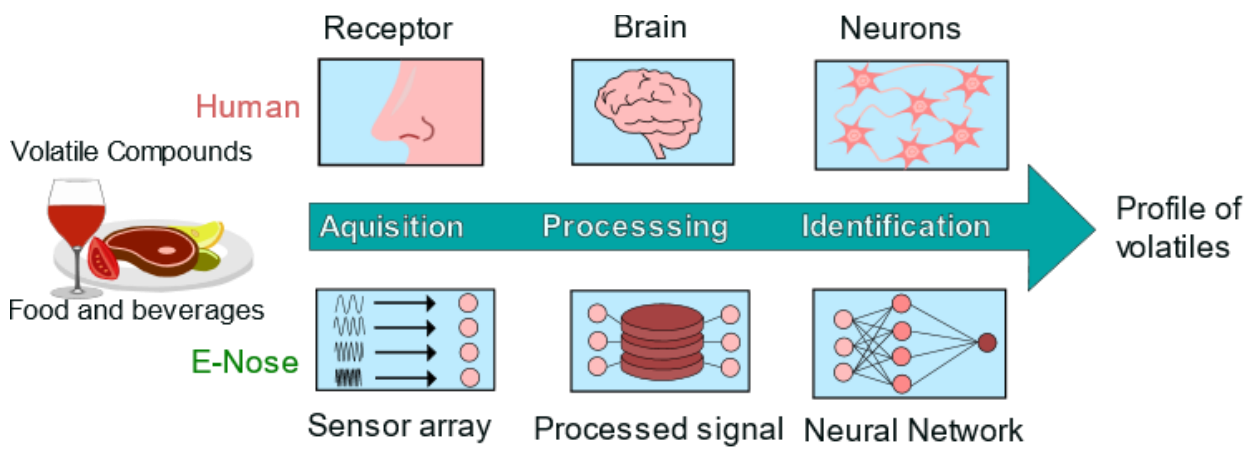

**Figure 4.** Parallel description of animal olfactory system and artificial E-nose system

Some of the commonly used commercial e-nose systems, such as the PEN3 developed by Airsense Analytics, The Cyranose® 320 produced by Sensingent, and the Fox (2000, 3000, 4000) instruments developed by Alpha-MOS, utilize electrochemical transduction. This involves converting the interaction between the sensitive layer and the target analyte into an electrical signal, which is then collected using analog or digital read-out devices. Therefore, this review will focus on electrochemical transduction as it provides reliable quantitative and qualitative detection and analysis.

An overview of the literature database revealed that significant efforts are being made to miniaturize the complete E-nose system in order to make such devices mobile and suitable for on-site applications. $MO_x$ are considered ideal candidates for miniaturization due to their high sensitivity per surface area. Technological advancements in fabrication of $MO_x$ layers enabled miniaturization to micro and nanoscale levels. Kang et al., 2020 have demonstrated a top-down lithography approach for fabrication of different $MO_x$ nanopattern channels (NiO, CuO, $Cr_2O_3$, $SnO_2$, and $WO_3$) on silicon wafer. Developed E-nose system was able to effectively differentiate seven hazardous analytes. Reported results offer a powerful tool for miniaturization of E-nose devices. Aiming to achieve low power consumption and RT operating conditions, Chen et al., 2018 developed an e-nose device with 3D $SnO_2$ nanotube array, and with the designed setup they achieved high-sensitivity and successful discrimination of gas mixtures at RT, which yielded 1000 times lower power consumption. Rehman et al., 2020 presented an interesting approach for wireless e-nose devices based on Figaro TGS sensor array, for discrimination of gas mixtures and VOCs.

Most significant advancement of $MO_x$ in the evolution of e-nose devices is expected to be found in the nexus between tailored material properties and further improvement of AI and machine learning algorithms for data analysis. In future endeavors, closer binding between E-nose and artificial intelligence technologies should be expected. Kiselev et al., 2018 published an interesting paper, describing functional instability of E-nose units over different time periods of exploitation. Through examination of the gathered data, authors discovered that subtle changes in the composition of the surrounding air had the greatest impact on the stability of device performance. The authors developed a novel strategy through an extra training technique that is shown to effectively control both the temporal changes in ambient and the drift of multisensor array characteristics, even over extended periods of time, to overcome the discovered instabilities. Within the suggested framework, using the power of AI algorithms, more efficiency of $MO_x$ materials can be extracted, opening a new frontier for scientific, technological and market development.

With the improvements in design of sensitive layers and pattern recognition systems, e-nose technology and devices expanded over the agricultural and food production fields. E-nose devices found application in monitoring of fruit, vegetables, grain products, dairy products, seafood products, fats, oils and beverages quality, authenticity, and other important parameters. Interesting application of the e-nose sensor array was reported by Ayari et al., 2018, where monitoring of the adulteration of margarine in cow ghee was performed. They used an array composed of commercial MQ sensors and TGS sensors, whereas for the data analysis they used principal component analysis (PCA) and artificial neural networks (ANN) methods. Based on the obtained results they were able to discriminate between the pure and counterfeit cow ghee samples. The freshness of post-harvest kiwifruit was predicted in the study by Du et al., 2019 using an E-nose with 10 $MO_x$ semiconductor (MOS) gas sensors. E-nose device has proven to be a powerful tool for the prediction of the postharvest kiwifruit ripeness through aroma volatiles. Trans-cinnamaldehyde, thymol, menthol, and vanillin were the four safe bioactive volatile terpenes and natural chemicals that were the subject of the study presented by Gouda et al., 2019 on the egg yolk volatile components. Gas chromatography/mass spectrometry (GC/MS) and headspace solid phase microextraction (HS-SPME) were used to examine the volatiles, and an E-nose with 18 sensors was used to distinguish between different fragrance patterns. To better comprehend the impacts of bio-active chemicals on the volatility of biological fatty media, their study revealed new information, describing how terpenes can shield egg yolk media from the unpleasant odor production brought on by the lipid and protein breakdown. Liu et al., 2018 used an E-nose to track the progress of fungal contamination in peaches. Three common spoilage fungus, Botrytis cinerea, Monilinia fructicola, and Rhizopus stolonifer, were injected into peaches before they were kept for different amounts of time. Then, e-nose was used to examine the volatile chemicals produced in the fungal-inoculated peaches, growth information (colony counts) of the fungi and the findings were cross-compared. The findings demonstrated a correlation between variations in volatile chemicals and the total number and kind of fungus present in fungal-inoculated peaches. The primary components of E-nose reactions were terpenes and aromatic chemicals. The outcomes also demonstrated the potential of e-nose for application in differentiating between types of fungal contamination in peaches by exhibiting its excellent discriminating accuracy. Liu et al., 2019b, reported an innovative technique for a bionic E-nose based on a MOS sensor array and machine learning algorithms utilized to identify wine quality. Odor recognition makes it easier to distinguish between wines with various characteristics, such as production regions, vintage years, fermentation techniques, and varietals. Results show the value of the created E-nose, which, after choosing the best algorithm, may be used to differentiate between various wines based on their characteristics. Pirsa and Shamusi, 2019 reported research regarding the viability of changing the e-nose software to assess rice aging during storage with a $MO_x$ gas sensor. Using an E-nose device, the scent change of aromatic and non-aromatic rice was tracked throughout the storage process. The aromatic samples followed a predetermined schedule with discrete groupings that stood out on their own, demonstrating the lowering of aging indices. During the early phases of preservation, the volatile components in the aromatic rice experienced significant modifications. Unorganized clustering of the non-aromatic rice demonstrated its resilience since the scent molecules it contains become less varied. The E-nose system coupled with advanced numerical techniques can be used to accurately manage rice aging.

The market for E-noses is expected to expand significantly in the next decade. Increasing usage of e-nose in the food and beverage sector, which aids in enhancing the quality of consumable items, is one of the reasons driving the growth of this market. The expansion of this market might be hampered by several issues, such as the relatively high price of e-nose devices. However, e-nose technology has seen growing application in a number of industries, including environmental monitoring and military defense, thanks to global players' innovation and continual development efforts. At the moment, e-nose and expansion of its functions in healthcare and quality control services have opened up new potential for this sector.

## 2.4 Smart packaging

The development of smart food packaging is increasingly important in modern lifestyle, playing a crucial role in reducing food waste (Motelica et al., 2020; Galstyan et al., 2018), reducing synthetic polymer waste through bio-based and biodegradable packaging (Salgado et al., 2021), and improving customer convenience in general. Smart packaging can protect food and extend its shelf life by modifying the atmosphere in a package, introducing antimicrobial or anti-oxidative properties, indicating spoilage, improving humidity and gas barrier properties, and monitoring storage conditions (Yousefi et al., 2019, Salgado et al., 2021).

For this purpose, NMs based on transition $MO_x$ exploit their antimicrobial and photocatalytic properties to protect packed products from the negative effects of various microorganisms and may provide additional beneficial effects on the shelf life of packed food. The most commonly used oxides for this purpose are $TiO_2$ and zinc dioxide ($ZnO_2$).

The mechanism of $TiO_2$ antimicrobial effect in nanocomposite films is explained in detail by Kubacka et al., 2014. It clearly indicates a great potential for this material to be implemented in smart food packaging, which was recognized by many researchers who employed different approaches toward development of $TiO_2$ based smart packaging. Recent research shows high antimicrobial activity of cellulose based paper composite with $TiO_2$. The results indicate outstanding antibacterial properties against escherichia coli (E. coli), moderate action against staphylococcus aureus (S. aureus), for composite paper with 8% of $TiO_2$ NPs, and good scalability of process at low production cost (Maślana et al., 2021). Active nanocomposite packaging based on $TiO_2$ and chitosan demonstrated better moisture barrier and mechanical properties than pure chitosan and efficient photo-degradation of ethylene for extended shelf-life (Kaewklin et al., 2018). Currently, an unconventional sort of smart packaging considers application of coating directly on a product, such as meat, cheese or a fruit. Fonseca et al. have developed gelatin-$TiO_2$ coated expanded polyethylene foam nets deposited on the surface of a papaya fruit by a novel procedure. After 4 days in storage, coated fruit have demonstrated lower ethylene emission and respiration rate at the climacteric peak, better physical properties, better preservation of green peel, orange pulp, sweetness/acidity equilibrium and no fungal growth (de Matos Fonseca et al., 2021). $TiO_2$ also finds an application in oxygen detection in the packages with modified atmosphere. The colorimetric printable sensor developed by Wei et al. based on $TiO_2$ nanotubes showed quick response by color change in the presence of oxygen, and its unaffected by the natural light (Wen et al., 2018). The oxygen sensor was composed of hydroxyethyl cellulose, glycerin, methylene blue and $TiO_2$ nanotubes, ball milled all together and deposited by screen-printing on a PET substrate. The indicator was UV activated before the use and showed a great contrast between referent and sensor exposed to oxygen in modified atmosphere packaging.

The potential of ZnO NPs has been recently reviewed by Kim et al., 2020. The review reports great potential of ZnO NM incorporated in smart food packaging, exhibiting strong biocidal properties and nontoxic behavior for humans. The review paper also explains the ZnO mechanism of photocatalytic effect and diversity of ZnO NP morphology. A great example of multifunctional ZnO based smart food packaging was developed by Pirsa and Shamusi, 2019. A newly developed cellulose-polypyrrole-ZnO film together with a chicken meat in a package was put under a test over time. The results showed a good antimicrobial and antioxidant protection of the meat product, prolonging good organoleptic properties. Also, electrical properties of the film had been related to storage conditions over time, making the film also an time-temperature indicator. A ZnO based bionanocomposite coating reported by Li et al., 2019a applied on bananas produced moderate protection of the fruit in a period of 7 days. The coating delayed ripening and improved overall postharvest quality and shelf life of the fruit. A suitable replacement for conventional synthetic polymer packages, based on zein-ZnO and ZnO: magnesium (Mg) QDs was reported by Schmitz et al., 2020. The composite films were transparent and homogenous. Hydrophobicity of the films was improved as well. An important feature of this active food package is seen in antimicrobial testing. Different amounts of ZnO fillers were tested, where best inhibition of *S. Aureus* was reported for sample with 44.8 wt% ZnO NPs in zein, while ZnO:Mg QDs were most efficient against *E. Coli* when less than 1 wt% of particles were added. ZnO NPs are used not only for antimicrobial purposes, but for improving mechanical properties as filler also. Chicken skin gelatin/tapioca starch composite with ZnO has been analyzed by Lee et al., 2020. It was found that tensile strength doubled with addition of 5% of ZnO, while elongation at break reduced. Water vapor permeability was also reduced, with a best result for the composite with 4% of ZnO. Antimicrobial properties were also tested against *E. Coli* and *S. Aureus*, showing good improvement with increasing content of ZnO up to 5%. The application of ZnO NPs as modifiers of some standard synthetic polymers, like low density polyethylene, was also explored. This approach looks to improve a packaging material in a simple procedure, adding an antimicrobial effect of the packaging (Rokbani et al., 2019). Different processing conditions were explored toward obtaining stable ZnO coated packaging material, without detachment of ZnO over time. It was concluded that even after 8 months of storage the material maintained moderate antimicrobial properties toward *E. Coli* and *S. Aureus*. It was also found that such coated material may lose antimicrobial properties after washing with water.

Semiconductor NMs based on $MO_x$ have made a breakthrough in food storage application as the so-called "photocatalytic refrigerators". The source of the idea is drawn from excellent photocatalytic properties of these materials, that can be used for antibacterial purposes, for odor elimination or for ethylene decomposition.

The incorporation of MOx into food packaging as an antimicrobial agent and nanosensors, as previously described, can transform standard packaging into advanced smart packaging. This technology not only helps preserve and maintain the freshness and quality of the food within, but also provides up-to-date information on the quality of the packaged food, enabling consumers to make informed decisions regarding purchase, storage, and consumption (Thirupathi Vasuki et al., 2023, Joshi et al., 2024).

Smart packaging aligns with the broader trend of incorporating technology into various aspects of daily life, providing consumers with more information and control over the products they consume. This innovation represents a step forward in creating a more sustainable, efficient, and consumer-friendly food supply chain.

While the potential benefits of $MO_x$-integrated packaging are substantial, ongoing research and collaboration across industries are essential to address any potential safety concerns. Adherence to regulatory guidelines is crucial to ensure that these technological advancements contribute positively to the food industry while maintaining consumer safety and confidence.

## 3. $MO_x$ NMs for agrochemical targeted delivery

Common agricultural practices mainly rely on the use of agrochemicals, such as fertilizers and pesticides, to maintain sufficient crop yield. Fertilizers act as plant growth enhancers by increasing nutrient availability and aiding mineral delivery, while pesticides are employed to protect plants from pests and weeds (Dhankhar and Kumar, 2023). In recent decades, the rapid progress of science, technology, and industry has led to the widespread integration of artificial NMs in agrochemical formulations. Within this class of materials, $MO_x$ NMs stand out due to their unique properties. As shown in Figure 5, NMs used to promote crop production can be classified into two categories: nanofertilizers and nanopesticides.

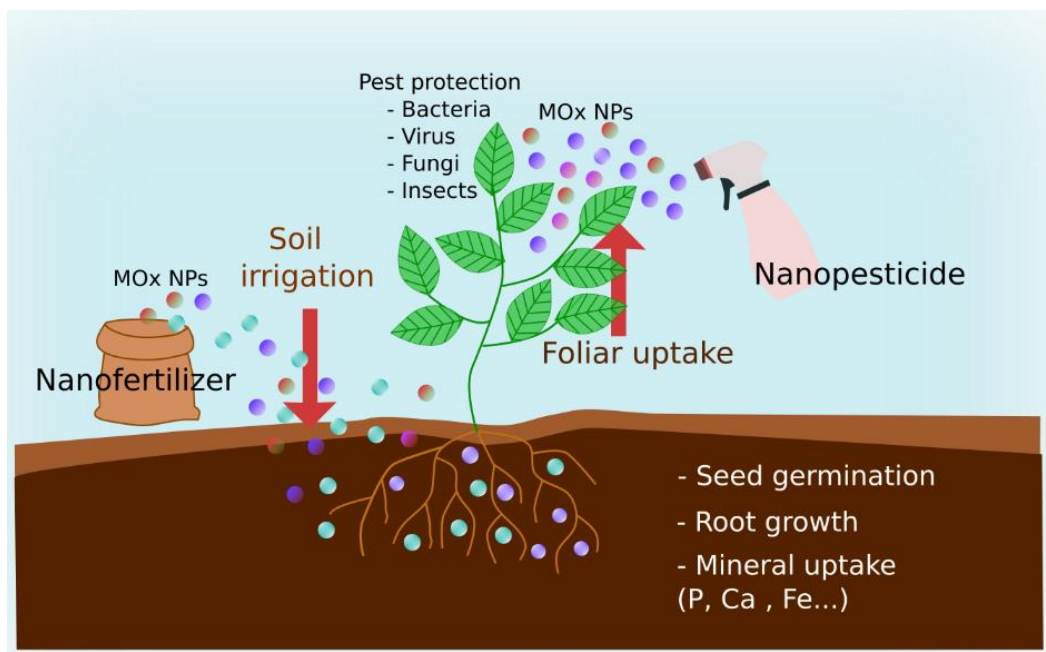

**Figure 5**. Schematic representation of multifunctionality of MO$_x$ NPs and their application as Nanofertilizers and Nanopesticides in agricultural environments.

The aim of this section is to bring such an important field of research closer to readers attention, and to provide more clarity in potential for further advancement of MO$_x$ NMs in crop production applications.

### 3.1. Nanofertilizers

Conventional use of fertilizing agents such as urea, nitrate, or phosphate compounds accumulates extensive amounts of harmful chemicals. The staggering demand for fertilizer application was forecast to significantly increase with a growing population, leading to increased efforts for efficient management of pesticides and fertilizers (Xiang et al., 2020). One proposed solution is to find alternative technologies for pesticide and fertilizer deployment. Another solution is directed towards the development of rapid pathogens and pest detection (Buja et al, 2021).

The introduction of NMs for the formulation of nanofertilizer has proven to be a promising solution to agricultural problems. Because of their outstanding characteristics such as their small size, large surface area to volume ratio and surface specificities they are able to achieve highly nutrient-dense foods that are balanced and highly nutritious. The advantages of using NM as a nanofertilizer are improved germination percentage by increasing their capacity to absorb water; enhancement in vegetative biomass; a higher photosynthetic rate, which then increases the light retention and light diffusion in the plant; a reduction in soil toxicity and a lessened effect of fertilizers taken in larger quantities on plants (Butt and Naseer, 2020). To achieve sustainable agriculture standards, NMs made a landmark impact due to their ability to slowly release fertilizers as well as localized targeted applications. In comparison to conventional fertilizers, nutrients in nanofertilizers are coated with a thin layer of NMs, which might be in the form of emulsions or NPs (DeRosa et al., 2010). Analysis of the interaction between nanofertilizers and plants revealed that it is not important how much of a nutrient is present in the soil, but how much of it is available. Nanofertilizer deployment can drastically reduce chemical waste in soils, fostering a cost-efficient and environmentally friendly economy. Nanofertilizers can be absorbed faster by the roots and can be assimilated easily into germinating seeds. Some of the most important benefits include controlled nutrient delivery, elevated seed germination, accelerated and enhanced plant growth, improved photosynthesis properties, increased chlorophyll formation, and stimulation of the rhizosphere and soil microflora. Highly intriguing research on how ZnO NPs morphology affects seed production & antioxidant defense in soybeans was published by Yusefi-Tanha et al., 2020. According to the research, the specific particle size and shape, along with a concentration of ZnO NPs had a substantial impact on the yield of soil-grown soybean seeds, lipid peroxidation, and several antioxidant indicators. In contrast to $Zn^{2+}$ ions, rod-shaped ZnO-NPs of 500 nm size, floral-shaped ZnO-NPs of 59 nm size,

& spherical-shaped ZnO-NPs of 38 nm, the spherical-shaped ZnO-NPs demonstrated the most shielding behavior, especially up to the concentration of 160 mg Zn/kg and also induce greater oxidative stress responses compared to other morphologically different ZnO-NPs, particularly at concentration of 400 mg Zn/kg, the spherical shape ZnO-NPs of 38 nm provide synergistic effects for soybean. According to this report, for improving crop yields, food quality, and reducing malnutrition globally, ZnO-NPs are high-potential candidates to be used as novel nanofertilizers for Zn-deficient soils.

A nanofertilizer incorporating ZnO and CuO NPs was reported by Amanda Ekanayake et al., 2021, which effectively releases micronutrients while nourishing the soil around it. The first nanonutrients utilized in this experiment were Zn and Cu, which help seeds germinate when administered as $MO_x$. Alginic acid was used to treat the soil. To make the fertilizing complex, which was made up of alginate chains that had been bridged with Zn (II) and Cu (II) and contained extra $MO_x$ NPs in a hydrogel, NPs were combined with sodium alginate. To investigate micronutrient absorption, tomato plants underwent a 30-day leaf examination. Nanofertilizer was mixed with prepared compost and utilized with manure on tomato plants to investigate the behavior of micronutrient discharge. The polymeric fertilizing complex described in this article was shown to have a high potential to progress into a multipurpose micronutrient nanofertilizer and to maintain the soil's organic condition.

Zahra et al., 2015 studied various significant parameters and naturally available soil-bound inorganic phosphorus (Pi) under the influence of $TiO_2$ NPs & iron oxide NPs ($Fe_3O_4$-NPs). *Lactuca sativa* (lettuce) was grown for three months to examine the effects of soil supplemented with $TiO_2$-NPs and $Fe_3O_4$-NPs, with increasing concentrations up to 250 mg/kg. To track translocation and comprehend potential processes for Pi uptake, a variety of spectroscopic and microscopic characterizations like Raman spectroscopy, scanning electron microscopy (SEM), Fourier transform infrared spectroscopy (FT-IR), etc. were used. For phosphorus buildup, different tendencies were detected, roots (have higher $TiO_2$-NPs greater than $Fe_3O_4$-NPs greater than control) and shoots ($Fe_3O_4$-NPs greater than $TiO_2$ - greater than control). The ability of $MO_x$ NPs to absorb phosphate ions, together with changes in phosphorus concentration and stress in the rhizosphere, might have contributed to increased root exudation and acidity. Phosphorus availability and plant uptake both increased as a result of each change.

Another study directed by Kolenčík et al., 2020 compared the effects of two nano-fertilizers, ZnO and $TiO_2$, on sunflower production. Despite the fact that $TiO_2$-NP plant enhancers exhibit greater photo-stability, it is crucial to acknowledge that their nutrient delivery is diminished, and they may have detrimental effects. Conversely, the ZnO-NPs that displayed lower resistance had a more pronounced impact on the physiological parameters of the sunflower. This can be attributable to Zn's increased bioavailability resulting from the transformation of ZnO-NPs. As a result of the ZnO-NPs treatment, sunflower physiological responses were induced, while the $TiO_2$-NPs treatment primarily affected oil content, which was increased by approximately 63.6% compared to 59.2% in the control and changed sunflower maturity early on. Furthermore, substantial disparities exist in the distribution of trichomes between the NPs-treatment techniques and the control, with particular emphasis on trichome width and length on the leaf surface.

Pt-functionalized iron oxide NPs and iron oxide at extreme (minimum and maximum) concentrations were used to study the effects of seed pre-soaking solutions on embryonic root development in legumes (Palchoudhury et al., 2018). This process uses less fertilizer, which means fewer NPs encounter the soil, making it more environmentally friendly. Using a combination of characterization techniques and statistical analysis, iron oxide NPs were demonstrated to promote root growth in a range of 88 to 366 percent for 5.54 to 103 mg/L of iron. At high concentrations of 27.7 mg/L, platinum-coated iron oxide NPs also reduced root growth. The undertaken approach, to precisely manage the design of necessary nanofertilizers for high-efficiency agriculture production may be applied to many environmental challenges.

To maximize nutrient utilization efficiency, Leonardi et al., 2021, have developed environmentally friendly hybrid nanocomposites composed of chitosan and sodium alginate that act as a biodegradable shell that encapsulates CuO NPs, as a nanofertilizer. Scanning and transmission electron microscopy (TEM) were used to determine the morphology characteristics of the samples, with the spherical shape of the nanocomposite measuring about 300 nm in diameter, while chemical analysis included X-ray diffraction (XRD), thermogravimetric analysis, FT-IR, and dynamic light scattering. As compared with bare CuO NPs, hybrid CuO/**polyelectrolyte complexes (PEC)** NPs exhibited 80% copper release after 22 days and showed efficacy in increasing *Fortunella margarita* Swingle seed germination.

Mahajan et al., 2011 examined the influence of nano-ZnO particles on the root and shoot growth of mungbean (*Vigna radiata*) and *Cicer arietinum* by introducing different concentrations of nano-ZnO particles in suspension of the agar media. An inductive coupled plasma/atomic emission spectroscopy (ICP-AES) method and scanning electron microscopy were employed to detect the absorption of NPs by seedling roots. Different concentrations of ZnO NPs have varying effects on the growth of mung and gram seedlings, with the most pronounced impact observed at 20 ppm for *Vigna radiata* and 1 ppm for gram (*Cicer arietinum*) seedlings. The growth of the bacteria was inhibited beyond this concentration. The aggregation and absorption of nano-ZnO particles by the roots may explain the successful growth at a particular optimal concentration and the reduced growth above this concentration.

calcium oxide (CaO) NPs offer an environmentally friendly option for fertilizers, given their abundance, minimal material usage, straightforward preparation, and low cost. In the presence of citric acid, a chelating agent, the direct influence of CaO NPs as nanofertilizers for foliar feeding was tested in the field for various fruits, resulting in an increase in the $Ca^{2+}$ content in the fruit peel and the cell wall width. Nonetheless, the amount to which plants absorb these NPs remains unknown, making it difficult to make conclusive statements about their possible consequences for human health (Carrasco-Correa et al., 2023).

ZnO NPs have demonstrated significant promise in the field of agricultural advancement, particularly when incorporated as a colloidal solution in nanofertilizers. This application has been observed to enhance crop growth and yield. Biogenic production is a crucial feature of ZnO NPs, as it is both eco-friendly and biocompatible. Additionally, their antimicrobial activity aids in safeguarding plants against microorganisms. The study conducted by Gauba et al., 2023 investigates the mechanism of uptake, translocation, and accumulation of ZnO NPs synthesized through green methods in plants. The study reveals the significant role of these NPs as a nanofertilizer in crop fields to enhance crop yield. Furthermore, it is imperative to investigate the potential correlation between the different attributes of NPs, including their surface area, size, and shape, and their efficacy in biological applications.

In comparison with silver (Ag) NP, ZnO NP exhibits lower toxicity to plants and could serve as a promising pesticide alternative (Chhipa, 2017). Using biogenic ZnO NPs, Keerthana et al, 2021 showed significantly improve growth parameters of *Abelmoschus esculentus L.*, such as seed germination (89 %), shoot height (18 cm), root height (7.8 cm), total plant height (34 cm), total fresh weight (22 gm), root fresh weight (1.8 gm), total dry weight (5.24 gm), leaf area (7.8 cm), shoot dry weight (4.76 gm), root dry weight (0.39 gm), number of branches (7), number of pod (7) and length of pod (13 cm). Table 2 reports some of the recent applications of nanofertilizer in agricultural plant production.

**TABLE 2. Different types of MOx nanofertilizers, their sizes, target plants, and following effects in agriculture.**

| Nanofertilizer | Size (nm) | Target plants/seeds | Effect | Reference |
|---|---|---|---|---|
| **CaO** | 25-40 nm | Various fruits | Increase in the $Ca^{2+}$ content in the fruit peel and the cell wall width | Carrasco-Correa et al., 2023 |
| **ZnO** | 38 nm | Soybean | Effect on seed yield, lipid peroxidation, and antioxidant biomarkers in soybean. | Yusefi-Tanha et al., 2020 |
| **ZnO** | 20 nm | Vigna radiata, Cicer arietinum | Effect on mung and gram seedling growth. | Mahajan et al., 2011 |
| **ZnO** | 24–28 nm | Mungbean cultivars | Higher expansion, yields, Zn content, antioxidative enzyme activity, and decreased accumulation. | Rani et al., 2023 |
| **ZnO** | 20-30 nm | Abelmoschus esculentus L. | Improved seed germination and seedling parameters. | Keerthana et al., 2021 |
| **ZnO** **Fe₃O₄** **AuSi** | 17.3 nm 5.4 nm - | Sunflowers | ZnO NPs improved the grain yield and phosphorus content. AuSi-NPs provided a greater amount of essential linoleic acid (54.37%). The $Fe_3O_4$ NPs treatment had beneficial effects on physiology, yield, and insect biodiversity. | Ernst et al., 2023 |
| **ZnO** **TiO₂** | 17.3 nm < 30 nm | Sunflower | $TiO_2$ improved sunflower quantitative and nutritional parameters, oil content (63.6%), and early plant maturation, ZnO better reflected the sunflower physiological parameters. | Kolenčík et al., 2020 |
| **ZnO** **CuO** | 80-100 nm 80-90 nm | Water, soil, tomato leaves | High potential to progress into a micronutrient nanofertilizer and maintain the soil's organic condition | Ekanayake et al., 2021 |
| **CuO/PEC** | 300 nm | Fortunella margarita Swingle seed | Increased seed germination. | Leonardi et al., 2021 |
| **MnOₓ/FeOₓ** | 23.42 nm | Corn | The development of plants, with a specific focus on germination rates, root growth, and fresh weight in maize plantlets. | de França Bettencourt et al., 2020 |
| **Fe₃O₄** | ~16 nm | Legume seeds | Increase of 88–366% of embryonic root growth in specific legume plants. | Palchoudhury et al., 2018 |
| **γ-Fe₂O₃ S-Fe₂O₃ S* small** | 40-215nm 4-15 nm | Soybeans | γ-Fe₂O₃ improved growth and decreased the Fe and N inputs. S-Fe₂O₃ enhanced the growth, yield, and nutritional quality, and provided better delivery of Fe nutrition with a slower rate of Fe dissolution. | Cao et al., 2022 |

### 3.2. Nanopesticides

NM-based formulations outperform traditional pesticides and formulations due to their high efficacy, which is due to their high surface area, enhanced solubility, induction of systemic activity resulting from smaller particle size, improved agility, and low toxicity for the removal of organic solvents (Sasson et al., 2007). NPs have the potential to significantly improve the effectiveness and stability of enzymes, entire cells and other natural products employed as biopesticides. The objective of utilizing NMs in agriculture is to mitigate nutrient scarcity, amplify crop production, minimize the reliance on chemical plant defense, and lower manufacturing expenses to optimize productivity (Hyder et al., 2023). While low-volume, high-value applications are anticipated for NM-based preparations, nanostructured delivery systems will necessitate a targeted distribution strategy centered on exploiting the pest's habits and behavior.

Jameel et al., 2020 created ZnO-NPs with a thiamethoxam nanocomposite and examined their synergistic effects on *Spodoptera litura larvae*. The larvae were fed castor leaves that had been treated with thiamethoxam as well as a mixture of ZnO-NPs and thiamethoxam (1090 mg/L). They found a 27% increase in larval mortality, as well as deformities in adults and pupae, delayed emergence, and reduced fecundity. Experiments on the pulse beetle, Callosobruchus maculatus *(C. maculatus)*, performed by Malaikozhundan and Vinodhini, revealed an attractive strategy using an environmentally friendly manufacturing process of Pongamia pinnata (*P. pinnata*) coated ZnO NPs (Pp-ZnO NPs). These NPs reduced the number of eggs deposited and the hatchability of *C. maculatus* in an inverse proportion to the dose. Following treatment with Pp-ZnO NPs reveals a substantial suppression in the larval, pupal, and overall development stages of *C. maculatus* was observed. Additionally, 25 g mL$^{-1}$of polypropylene (Pp)-ZnO NPs are 100% fatal to *C. maculatus*, making them more effective in controlling the species. Pp-ZnO NPs are more competent in suppressing *C. maculatus*, inducing 100% mortality at a concentration of 25 g mL$^{-1}$. The midgut activities of glutathione S-transferase, lipase, β-amylase, cysteine protease, and glucosidase in *C. maculatus* were reduced post-stimulation with Pp-ZnO NPs (Malaikozhundan and Vinodhini, 2018).

The objective of the research presented by Ilkhechi et al., 2021 was to employ the sol-gel method to create ZnO, TiO$_2$, and ZnO-TiO$_2$ NPs with a weight ratio of 1 to 1 for Zn and Ti utilizing zinc acetate and titanium isopropoxide. Aspergillus flavus (*A. flavus*) was tested in vitro using antifungal activities such as minimum inhibitory concentration (MIC) and minimum fungicidal concentration (MFC) to assess NPs' morphological and physical properties. Although ZnO-TiO$_2$ was more efficient against *A. flavus* than pristine TiO$_2$ and ZnO, all the produced NPs at (50 g/ml) concentration inhibited fungal growth. Spurs growth was completely suppressed by TiO$_2$ and ZnO-TiO$_2$ for 300 g/ml) concentration. Pure ZnO and TiO$_2$ had pyramidal and spherical shape, respectively, whereas ZnO-TiO$_2$ NPs had both spherical and pyramidal shapes on the surface with growing particles. A modest concentration (150 g/ml) of ZnO-TiO$_2$ revealed increased reactive oxygen species (ROS) formation and oxidative stress induction as compared to TiO$_2$ and ZnO alone, leading to the fungicide's fungicide activity (Ilkhechi et al., 2021). In conclusion, the nanostructured ZnO-TiO$_2$ composite can be employed as an antifungal medicine, however more study is required to comprehend how the NPs' antifungal mechanism differs from ROS-induced apoptosis.

In their work, Kolenčík et al., 2019 discovered that ZnO NPs applied foliarly improve the quality of crop products, as the plant water stress index was lower after ZnO NP foliar application than in the control throughout the entire life cycle. According to the increased nutritional parameters of the foxtail millet grain, the plant's photosynthetic efficiency, transpiration, and enzyme activity have improved. By incorporating ZnO NPs, the oil content was enhanced by 34% and total proteins by 9.1%. Additionally, the seed yield and seed Zn accumulation exhibited a significant increase of 18.3% and 21.1%, respectively, compared to Bulk-Zinc sulfate (ZnSO$_4$). On the other hand, no differences were observed between ZnO NPs foliarly applied treatments and the control in terms of the weights of dry seed heads and thousand grains of ZnO NPs. The reason behind this was the considerable decrease in zinc concentration, falling below the typically referenced 0.1% level that is known to have detrimental effects when Zn$^{2+}$-corresponding metal-based fertilizers are applied to the leaves.

Lakshmeesha et al., 2020, created high-purity, nano-sized (30-40 nm) hexagonal-shaped ZnO-NPs from aqueous leaf extract of *Melia azedarach* (MaZnO-NPs) in order to reduce antifungal activity. Characterization of the generated MaZnO-NPs was accomplished using UV-Vis spectroscopy, FT-IR, XRD, SEM, and TEM. The nanocomposite was shown to be far more powerful than the synthetic fungicidal drug Amphotericin B, as evidenced by dose-dependent inhibition of cladosporium cladosporioides (*C. cladosporioides)* and Fusarium oxysporum (*F. oxysporum)* growth. MIC and MFC values for MaZnO-NPs nanocomposite were 81.67 and 178.3 mg/mL, and 93.33 and 208.3 mg/mL, while synthetic fungicidal agent Amphotericin B exhibited a MIC and MFC values of 203.3 and 326.7 mg/mL and 236.7 and 381.7 mg/mL against *C. cladosporioides* and *F. oxysporum*, respectively.

Malaikozhundan and Vinodhini, 2018 presented a green synthesis of *P. pinnata* leaf extract coated ZnO NPs (Pp-ZnO NPs) as an insect pest control agent. X-ray spectroscopy revealed hexagonal wurtzite structures in Pp-ZnO NPs with particle diameters of 21.3 nm. Pp-ZnO NPs were investigated for pesticidal activity against the pulse beetle, C. maculatus and shown to lower fecundity (eggs laid), pupal, and total development duration in a dose-dependent manner. At 25 g/mL, nanocomposite caused 100% mortality, while decreasing mid-gut α-amylase, α-glucosidase, β-glucosidase, cysteine protease, glutathione S-transferase, and lipase activity in C. maculatus. Manzoor et al., 2023 synthesized CuO NPs from A. indica leaves via green synthesis. CuO NPs with sizes ranging from 20 to 80 nm were discovered using TEM and SEM pictures. Synthesized CuO NPs demonstrated strong antibacterial activities against microorganisms, inhibiting pathogen activity by breaking cell walls and inhibiting fungal spore generation on jujube fruit. As a result of CuO NPs, ROS are produced, DNA is disrupted, and bacterial proteins are denatured, all of which inhibit the growth of bacteria. The approach described may be well-suited for applications in the biomedical, pharmaceutical, and agricultural fields. Table 3 reports some of the recent applications of nanopesticide based on $MO_x$ in plant protection.

**TABLE 3. Different types of $MO_x$ nanopesticides, their sizes, target, and following effects in plant protection.**

| Nanopesticide | Size (nm) | Target | Effect | Reference |
|---|---|---|---|---|

| Thiamethoxam/ZnO-NPs | 5 and 0.5 μm | *Spodoptera litura larvae* | Increase of 27% in larval mortality, deformities in full-grown and pupae, abeyant emergence, and attenuated fecundity. | Jameel et al., 2020 |
|---|---|---|---|---|
| Pp-ZnO NPs (P. pinnata coated ZnO NPs) | 21.3 nm | *Callosobruchus maculatus* | Delayed the overall development and suppression of *C. maculatus*. | Malaikozh undan and Vinodhini, 2018 |
| ZnO NPs, $TiO_2$ NPs, $ZnO-TiO_2$ NPs, | 33.21 nm 17.68 nm ZnO = 19.25 nm $TiO_2$ = 8.36 nm | *Aspergillus flavus* | Suppressed fungal growth. | Ilkhechi et al., 2021 |
| ZnO NPs | 20 nm | *Setaria italica L.* (foxtail millet grains) | Plants revealed inequalities in millet grain oil and total nitrogen content nutritional parameters. | Kolenčík et al., 2019 |
| MaZnO-NPs | 30-40 nm | Soybean | Inhibited the growth of *C. cladosporioides* and *F. oxysporum.* | Lakshmeesha et al., 2020 |
| $Co_3O_4$NPs | 34.9 nm | Rice plants | Inhibition of growth and biofilm formation of Xanthomonas oryzae pv. oryzae. | Ogunyemi et al., 2023 |
| CuO NPs | - | Wheat (*Triticum aestivum*) | Inhibition of root growth. | Xu et al., 2023 |
| CuO NPs | 29 - 45 nm | *Spodoptera frugiperda* | High larvicidal and antifeedant activity - reduced the number of larval hemocytes. | Rahman et al., 2022 |
| CuO NPs | 20 - 80 nm | *Ziziphus jujuba* fruit | Antibacterial, antifungal, and antioxidant properties. | Manzoor et al., 2023 |
| ZnO NPs | 20–30 nm | Wheat | Reduced Cd accretion in wheat grains. | Rizwan et al., 2019 |

In summary, $MO_x$ NPs have a high surface area-to-volume ratio, which enables them to disperse and be absorbed more effectively in plants. This improves the delivery of nutrients and helps in controlling pests. The small size of these nanoparticles also makes it easier for them to penetrate plant cells, thereby enhancing nutrient absorption and promoting growth. As a result, there is an increase in crop yields and the quality of produce is improved. On the other hand, nanofertilizers make use of $MO_x$ NPs to deliver nutrients directly to plant roots, enhancing nutrient uptake and utilization. This leads to stronger and healthier plants that are more resistant to diseases and environmental stress. However, there is still a lot to explore in this research field, and several promising directions for future investigation can be identified. Among these is the potential of MOx NMs for mitigating the negative effects of drought stress on plant growth. Certain nanoparticles such as ZnO, Si NPs can improve water retention in soil, promote water uptake by plants, and reduce the impact of drought on crop yield and may also be used to develop smart delivery systems for water and nutrients to plants during periods of water scarcity (Raza et al., 2023, Muhammad et al., 2022) . Research in this area aims to develop innovative solutions for sustainable agriculture, mitigating the impact of drought on crop yields.

## 4. NMs for agricultural environment

### 4.1 Agricultural wastewater treatment

Nowadays, the number and concentration of different pesticides in surface, ground and wastewater are significantly increasing, because of the great development of the agrochemical industry (Radović et al., 2015; Barbosa et al., 2016). Many of them do not decompose into simpler, less dangerous compounds, but accumulate in the environment and transform into even more dangerous forms (Reddy and Kim, 2015). The reason for this behavior is the fact that most pesticides are resistant to chemical and natural photochemical and biological degradations (Ajiboye et al., 2020). Keeping in mind all the above, the creation and implementation of effective techniques for the treatment of polluted water are crucial jobs. Various processes have been tested to reduce the concentration and potential health risk of pesticides (Reddy and Kim, 2015). Traditional methods (adsorption, nanofiltration, reverse osmosis, chemical treatments, and various biological processes) have several disadvantages, e.g. adsorption transfers the pollutant only from one phase to another, chemical oxidation can lead to incomplete degradation of pesticides, and in the case of biological treatments the main disadvantages are poor reaction rates, sludge removal and the need for strict control of suitable pH and temperature (Dong et al., 2015; Ong et al., 2018). Therefore, great attention is paid to Advanced Oxidation Processes (AOPs) that are employed for purification of water systems contaminated with pesticides. Among different AOPs, the central place is occupied by the processes of heterogeneous photocatalysis with the use of $MO_x$ NMs as photocatalysts. Namely, photocatalysts participate in and accelerate the chemical transformation of the substrate, but remain unchanged at the end of the photocatalytic process. Most photocatalysts are semiconductors, which are characterized by a suitable bangap (Dalrymple et al., 2007). Figure 6 shows an overview of publications for the last six years, which are based on the application of NMs in the removal of pesticides. In general, it is evident that there are more publications on this subject. Different semiconductor materials are used as photocatalysts in heterogeneous photocatalysis such as $TiO_2$, ZnO, $SnO_2$, ZnS, $WO_3$, cadmium sulfide (CdS), cadmium selenide (CdSe), gallium arsenide (GaAs), gallium phosphide (GaP), etc. and among all mentioned, $TiO_2$ and ZnO are the most frequently used ones (Etacheri et al., 2015; Ribeiro et al., 2015).

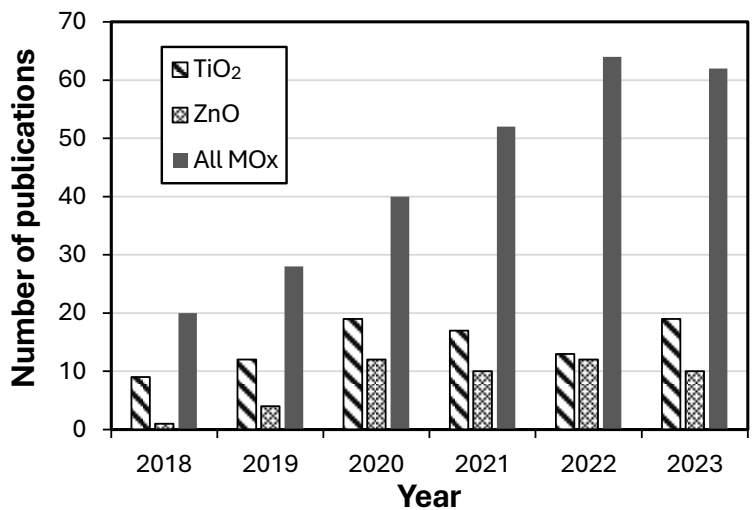

**Figure 6.** Number of articles published in 2018–2023, related to application of NMs in pesticides removal with "photocatal*", "removal", "ZnO", "TiO₂", and "pesticide*" as a searching keywords (available online on Scopus).

$TiO_2$ has been the most studied photocatalyst because of its low production cost and chemical stability as well as its ability to induce reductive and oxidative reactions (Dong et al., 2015; Ong et al., 2018). According to published works in the last period, special emphasis is placed on the modification and immobilization of $TiO_2$ to improve pesticide removal efficiency. Namely, Yu et al., 2015 investigated removal of pentachlorophenol using a photocatalyst made of Ag NPs mounted on anatase $TiO_2$ nanotubes. Their findings point that $Ag/TiO_2$ showed significant visible-light absorption which indicates the possibility for the sunlight driven photocatalytic degradation of pentachlorophenol. The enhancement of photocatalytic activity can be attributed to the strong localized surface plasmon resonance of the Ag NPs and the intimate contact between Ag NPs and anatase $TiO_2$ nanotubes, which favors the separation of photo-generated charge carriers. Through sol–gel synthesis, Achamo and Yadav, 2016 prepared nanosize Ag–N–P–tridoped $TiO_2$. A synergetic effect of tridoping $TiO_2$ enhanced its photocatalytic activity and thereby enhanced its ability to remove 4-nitrophenol from aqueous solutions. Obtained degradation of 4-nitrophenol using Ag–N–P tridoped can be explained by two effects, minimization of electron–hole recombination by the doped Ag and extension of the photoabsorption in the visible region by doped N and P. Also, An et al., 2016 investigated carbon-doped $TiO_2$ catalysts supported by zeolite in the removal of 18 pesticides and pharmaceuticals and their findings showed that $TiO_2$ can effectively degrade investigated substrates in water. Photodegradation efficiency is a consequence of carbon-doped $TiO_2$ loading on zeolites which could block their micropores and reduce their surface areas and micropore number. Besides, the activity of carbon-doped $TiO_2$ depends also on the photodegradable characteristics of the pollutants, since the zeolite support connects photolytic pollutants and carbon-doped $TiO_2$ for energy and/or electron transfer. Further, Khavar et al., 2018 investigated synthesis by ultrasonic-hydrothermal technique of In,S co-doped $TiO_2$ with reduced graphene oxide and its application in the removal of atrazine. Results showed complete degradation and 95.5% mineralization of atrazine within only 20 min. The enhanced photocatalytic activity of the prepared NM is explained by the synergistic effect of dopants contributing to improved visible light absorption and decreased recombination rate of the charge carriers. Besides, Abdelhaleem and Chu, 2019 studied application of a hybrid process for degradation of CBF through peroxymonosulfate activation and Fe(III) impregnated N-doped $TiO_2$ photocatalyst. Applied process exhibited about 90% of total organic carbon reduction which can be ascribed to the generation of more reactive radicals through peroxymonosulfate activation by Fe (III). The efficiency of $TiO_2$ for the degradation of mesotrione has also been improved by Merkulov et al., 2020 by using different Au NPs. Mesotrione can be efficiently eliminated from water by the modified $TiO_2$ and the reason for this catalytic performance might be the band gap energy, as in the case of Au modified $TiO_2$ it is shifted towards the lower values, hence there was efficacious use of visible light. Finčur et al., 2021 studied the efficiency of $TiO_2$, ZnO, and MgO (prepared by sol–gel method) in degradation of two antibiotics and two herbicides exposed to UV/simulated sunlight. Obtained results showed that $TiO_2$ proved to be the most efficient nanopowder under both irradiations. Besides, Ivetić et al., 2021 investigated the removal efficiency of organic water pollutants (pesticides quinmerac and tembotrione, and pharmaceuticals metoprolol, amitriptyline, ciprofloxacin, and ceftriaxone) from water by adsorption and photocatalysis using titanium/molybdenum/mixed-oxides. Findings indicated that for all substrates, UV light had a better removal efficiency than simulated sunlight, which can be correlated to UV-Vis reflectivity results and estimated values of the catalysts' optical absorption thresholds that just reach the Vis region, so when the reaction system is exposed to simulated sunlight, there is a smaller number of photons from the UV part of the spectrum, and thus, a smaller number of highly reactive species formed. However, an excellent outcome was observed for ciprofloxacin eradication (80%) by employing a synergic adsorption/photocatalytic method.

Many attempts are aimed at immobilization of $TiO_2$ on appropriate carriers. Assalin et al., 2016 describes application of immobilized $TiO_2$ using a flat panel photoreactor in the treatment of agricultural waste containing methyl parathion. Results showed that almost total mineralization of the insecticide was achieved after 90 min of the process. Khan et al., 2017 published the findings of the photocatalytic performance of a sol-gel nanostructured S-doped $TiO_2$ film for lindane removal with addition of peroxymonosulfate. Obtained results showed that addition of peroxymonosulfate significantly improved lindane removal from water which was potentially due to the dual role of peroxymonosulfate (as electron acceptor, thereby reducing the rate of electron-hole recombination and as an efficient source of sulfate anion and hydroxyl radicals). Fiorenza et al., 2020 reported that the 2,4-dichlorophenoxyacetic acid and imidacloprid photodegradation employing molecularly imprinted $TiO_2$ catalysts produced using the sol-gel methodology. Remarkable enhancement of the photocatalytic activity with the imprinted $TiO_2$ was established. Besides, ZnO has emerged as an effective and promising contender in the green environmental protection system (Lee et al., 2016; Ong et al., 2018). As in the case of $TiO_2$, special attention is paid to the modification and immobilization of ZnO. Namely, Shirzad-Siboni et al., 2017 prepared Cu-doped ZnO nanorods via a straightforward co-precipitation technique and investigated their efficiency in photocatalytic degradation of diazinon. Obtained results showed that the removal process using nanorods is more efficient as compared to UV/ZnO in degrading diazinon due to its low electron-hole recombination rate. Therewith, Hossaini et al., 2017 examined the effectiveness of synthesized C, N, and S-doped ZnO in degradation and mineralization of the same pesticide, diazinon. Doping ZnO with nonmetals enhanced the photocatalytic activity and mineralization of diazinon possibly due to the ion implantation in the bulk and surface of the ZnO which reduced the defect sites, improved the charge transmission and introduced some midstates in the bond gap of the ZnO. Besides, Kumari et al., 2020 synthesized ZnO/cerium dioxide ($CeO_2$) hybrid nanostructures via hydrothermal method. The photocatalytic efficiency is measured using rhodamine B and triclopyr. Results showed improved photocatalytic activity of $ZnO/CeO_2$ as a consequence of the prevention of the electron–hole pairs recombination. Maleki et al., 2020 studied the efficiency of $WO_3$ doped ZnO in the removal of diazinon. According to the obtained results, the addition of $WO_3$ to the ZnO increased ZnO efficiency in the photocatalytic process as a consequence of the network constant  reduction and increased ZnO density, which in turn improved its efficacy in the photocatalytic process. One very interesting aspect of NM preparation is biosynthesis. In work Pathania et al., 2021 synthesis of Cu-doped ZnO was performed using plant extract. This material was tested in photocatalytic degradation of chlorpyrifos and obtained results showed high activity in the degradation process by preventing the fast recombination of electrons and holes on the surface of the catalyst. Also, Rani et al., 2021 have synthesized ZnO coupled cadmium sulfide (CdS) nanocomposite using green synthesis method which included aqueous leaf extract of *Azadirachta indica*. The efficiency of prepared nanocomposite as photocatalyst was evaluated in degradation of chlorpyrifos and atrazine under sunlight. The coupled nanocomposite with high surface area and more negative zeta potential was able to adsorb a higher amount of pesticides. Besides, high photoactivity was supported by a large charge separation followed by the production of hydroxyl radicals for the degradation of investigated pesticides under direct sunlight.

Special attention is paid to testing of different nanocomposite materials since these materials can provide a superior photocatalytic performance owing to an extended light absorption range and an effective charge transfer. The work done by Jonidi-Jafaria et al., 2015 highlights the efficiency of ZnO–TiO$_2$ composite as a photocatalyst for diazinon degradation. Also, Merkulov et al., 2018 carried out a thorough investigation into the removal of certain pharmaceuticals and pesticides from various water types in the presence of manufactured bare TiO$_2$ and TiO$_2$/PANI nanocomposite powders. Furthermore, Boruah and Das, 2020 reported application of magnetic Fe$_3$O$_4$-TiO$_2$/reduced graphene oxide nanocomposite as an artificial nanozyme for the colorimetric detection and photodegradation of atrazine in an aqueous medium. Also, Luna-Sanguino et al., 2020 prepared TiO$_2$-reduced graphene oxide nanocomposites by hydrothermal method and investigated their catalytic efficiency on the pilot plant scale solar assisted photodegradation of the mixture of four pesticides. Ebrahimi et al., 2020 studied the efficiency of photocatalytic degradation of 2,4-dichlorophenoxyacetic acid using synthesized manganese (Mn)-doped ZnO/graphene nanocomposite. Furthermore, in work Anirudhan et al., 2021 synthesized a composite made of magnetically separable cysteine-modified graphene oxide, nickel ferrite, and $TiO_2$ was developed to remove the pesticide chlorpyrifos from aqueous solutions and then to photodegrade it.

All studies indicate the importance of the application of $MO_x$ NMs in the processes of removing pesticides from the environment. Regardless of that there are numerous publications on this subject, there is still a tendency to find a more efficient photocatalyst in terms of the pollutant removal, applicability in real conditions, as well as in the economic point of view.

### 3.2. Soil remediation

The focus of transdisciplinary research is on using engineered NMs to clean up contaminated soils. The phytoavailability of pollutants (As, Pb), nutrients (N, P), and trace elements (Cu, Fe, Mn, Zn) was regularly monitored in three Australian soils exposed to As or Pb pollution over a span of less than ten-year period in this research by Duncan and Owens, 2019. Soils were exposed with $nTiO_2$ or $nCeO_2$ NPs at a concentration of 500 mg $kg^{-1}$. Phytoavailability of contaminants, nutrients, and trace elements was evaluated for 260 days. Pb phytoavailability was enhanced by almost a factor of two in certain soils when $nCeO_2$ was present at the end of the experiment (day 260), which was explained by the fact that $nCeO_2$ lowers soil pH and/or competes with $Pb^{2+}$ ions for sorption sites while as phytoavailability was not affected.

Due to its reliability, cost, and overall environmental safety, the immobilization/adsorption methodology has emerged as the most prominent remedial strategy for eliminating pollutants from metal-contaminated soils. A crucial phase in the immobilization process is identifying the right materials and/or coating reagents based on the distinct contaminants and soil characteristics. The surface morphology of $MO_x$ NPs and their nanocomposites make them suitable candidates as adsorbing agents. Among various $MO_x$ NPs, $Fe_3O_4$ and $TiO_2$ NPs are explored extensively, for adsorbing various pollutants from the soil (Zeshan et al., 2022) . The $Fe_3O_4$ NMs have the capability to absorb and immobilize heavy metals like Cd, Pb and As (Wu et al., 2023 ; Wang et al., 2022b, Qian et al., 2020). $Fe_3O_4$ had the highest Cd sorption effectiveness at optimized condition with pH 6.0 and temperature between 10-20°C, showed highest adsorption capacity of 37.03 mg $g^{-1}$ as shown by Sebastian et al., 2019. Konate et al., 2017 studied the influence of $Fe_3O_4$ NPs (nano- $Fe_3O_4$) reduction of heavy metal toxicity (Pb, Zn, Cd, Cu) in wheat seedlings, which correlates with adsorption capability of nano- $Fe_3O_4$ for heavy metals. The particle sizes ranged from around 3.70 nm to 12.10 nm, according to the TEM picture. The addition of $Fe_3O_4$ NPs (2000 mg/L) in each heavy metal solution (1 mM) dramatically reduced the growth inhibition and oxidative stress in wheat seedlings.

In the study conducted by Dong et al., 2018, an innovative and economical method for remediating marine sediments polluted with polycyclic aromatic hydrocarbons (PAHs) was presented. The technique involves the use of iron oxide ($Fe_3O_4$) NPs supported on carbon black (CB) as a catalyst. The assessment of PAHs elimination was conducted using $Fe_3O_4$, CB, and $Fe_3O_4$/CB, which yielded removal rates of 75%, 64%, and 86%, respectively, at a temperature of 303 K with a concentration of 0.01 g/L. Majumdar et al., 2023 employed rice straw to make rice charcoal (RBC) doped with $Fe_3O_4$ NPs as eco-friendly composites. RBC was employed in the experiment at three different doses and coupled with $Fe_3O_4$ NPs. The use of these remedies reduced soil As bioavailability by 43.9%, 60.5%, and 57.3% for 0.5%, 1%, and 1.5%, respectively. As was adsorbed onto the RBC + iodine monoxide (IO) conjugate at a rate of around 3.28%, compared to 3.18% in RBC alone. Compared with control plants, rice seedlings treated with 1% RBC + IO conjugate were more stress-tolerant and generated fewer antioxidant enzymes. Mirzaee and Sartaj, 2023 investigated the efficiency of soil washing in conjunction with adsorption using magnetized granular activated carbon (MGAC) to remove polycyclic aromatic hydrocarbons (PAHs) from contaminated soil samples. The GAC/$Fe_3O_4$ composite was created using a co-precipitation process, and magnetite NPs were validated using XRD, FT-IR, and energy-dispersive X-Ray (EDS) methods. According to the results of soil washing, the highest removal efficiency was 45.8, 48.0%, and 47.4% for low molecular weight (LMW), high molecular weight (HMW), and total PAHs. Approximately 95% of the PAH-loaded MGAC particles were recovered from soil after treatment, suggesting that only a small amount remained in the soil, which should not cause significant contamination.

Wang et al., 2022c revealed that $MoS_2$ nanosheets with high Hg absorption capacity, rapid elimination kinetics, and superior selectivity may efficiently remove Hg in groundwater. $MoS_2$ nanosheets offer considerable promise in the treatment of Hg-contaminated groundwater due to their outstanding dispersity, low toxicity, and stability. Because of the high affinity between Hg(II) and $MoS_2$ via strong Lewis interactions, surface adsorption is the primary removal route for Hg in groundwater. As a result of the strong attraction between $MoS_2$ and Hg ions through Lewis soft-soft interactions, the preferential removal of Hg ions from deionized water occurred, with an anticipated binding to multiple sulfur atoms on the $MoS_2$ surface. The presence of chlorine in groundwater leads to an increase in the production of HgClOH, enhancing both the kinetic and thermodynamic reduction of Hg(II) to $Hg_2Cl_2$ and HgO. This process becomes the primary mechanism for Hg elimination that significantly enhances the overall capacity of $MoS_2$ nanosheets, resulting in an impressive removal ability of 6288 mg $g^{-1}$. This is one of the highest reported capacities for Hg removal thus far.

Hussain et al., 2018 investigated the effects of ZnO NPs on wheat (Triticum aestivum) growth, yield, antioxidant enzymes, Cd and Zn concentrations, and discovered that ZnO NPs were capable of lowering the toxicity and concentration of Cd in wheat by 30-77% and 16-78% with foliar and soil application, respectively. ZnO NPs improved growth, photosynthesis, and grain production, whereas Zn increased in wheat shoots, roots, and grains. Furthermore, ZnO NPs decreased electrolyte leakage while enhancing superoxide dismutase (SOD) and peroxidase (POD) activities in wheat leaves. SOD activity increased by 40% and 46% in the highest treatment (100 ppm), whereas POD activity increased by 46 and 54% in foliar and soil applications, respectively.

The objective of the research carried out by Emamverdian et al., 2020 was to assess the effects of $SiO_2$ NPs on the growth, photosynthesis, and protective enzymes of Arundinaria pygmaea, a bamboo species, under three different heavy metal stresses (Cu, Mn, Cd). According to the findings, the application of $SiO_2$ NPs in the presence of heavy metals (Cu and Mn) led to a considerable increase in plant growth compared to the control group. Moreover, these NPs were observed to alleviate the toxicity caused by Cu and Mn, as evidenced by significant improvements in protective enzyme activity, chlorophyll content, fluorescence, and plant biomass. Regardless, the administration of different treatments incorporating Cd concentrations led to a considerable decrease in biomass and shoot length. This finding suggests that the treatment with 100 μM $SiO_2$ NPs did not effectively hinder the accumulation of Cd in leaves. The research established that the utilization of 100 μM $SiO_2$ NPs improved the tolerance of bamboo species to heavy metals stress (Cu and Mn), thereby promoting plant growth. This improvement was attributed to the inhibitory effect of the NPs on metal accumulation in the leaves through the adsorption of metal ions.

Shaik et al., 2020 investigate the impact of green-synthesized ZnO NPs on chromium reduction activity, germination, and growth promotion of fenugreek seeds utilizing an aqueous root extract of *Sphagneticola trilobata* Lin. SEM pictures revealed the presence of zinc NPs in the nano range, with NP sizes ranging from 65 to 80 nm. As a result of NP doses of 1, 0.5, 0.25, and 0.10 g/L, the

% chromium metal reduction was 81.17%, 55.83%, 53.33%, and 38.17%, respectively. In comparison with control and zinc sulfate treatments, zinc NPs resulted in the highest seed germination, root and plant increase.

In their study, Song et al., 2019 showed that 100 mg/kg of $SiO_2$, $TiO_2$, ZnS, and $MoS_2$ NMs had the capacity to minimize hazardous heavy metal buildup in cucumber plants. The findings revealed that $MoS_2$ NMs influenced the accumulation of the majority of heavy metals. However, NMs with a large surface area and binding sites may also combine with cations such as $K^+$, $Ca^{2+}$, and $Fe^{3+}$, lowering plant absorption of macro- and micronutrients, which has a detrimental influence on plant development. Cucumber leaves' Si concentration rose in the presence of $MoS_2$, possibly promoting resistance to attacks. During a 4-week incubation period, none of the evaluated NMs had an effect on cucumber plant biomass.

## 5. Market barriers and Risk assessment

$MO_x$ NMs in the agri-food chain bring tremendous opportunities for both industry and consumers. However, there are several obstacles that need to be considered in order for NM technology to be adopted in the agri-food industry. These barriers encompass various aspects that are interconnected, including concerns surrounding nanotoxicity, safety, and regulatory harmonization.

Although extensive research has been conducted on $MO_x$ NPs such as ZnO and $TiO_2$, there have been conflicting findings regarding their potential toxicity. Significant uncertainty remains regarding their impact on human health and the environment, necessitating a clear risk assessment strategy (Sengul et al., 2020). The contamination of food products with NPs may occur at different stages of the food supply chain, starting from the use of NPs in pesticides for crop production and water purification techniques, to their involvement in nanosensing or packaging, which could result in their migration into food products.

The foundation for evaluating the risk of $MO_x$ NMs used in agriculture and food packaging lies in conducting physicochemical characterization. The safety of NMs is influenced by factors such as size, shape, surface charge, surface area, purity, stability, concentration, and agglomeration state, which can vary throughout their life cycle across the entire food supply chain. Furthermore, it is essential to assess the "exposure risk" posed by NMs. These materials can enter the human body through three different pathways: inhalation (respiratory tracts), ingestion (digestive tracts), and skin contact (blood vessels). In the case of incorporating $MO_x$ in packaging, it is crucial to evaluate the potential migration of nanostructures into the food. Lastly, it is important to identify and characterize specific hazards such as genotoxicity and cytotoxicity (Buzea et al., 2007).

Since nanotechnology was introduced in the agrifood sector, efforts were made worldwide to regulate the production and use of NMs through legislation, recommendations, and guidelines. For example, in the US, guidelines were established for risk assessments of NMs, such as the Final Guidance for Industry issued by the Food and Drug Administration (FDA) in 2014. In Europe, the European Food Safety Authority (EFSA) has developed a new guidance for the risk assessment of NMs in various food-related areas, covering everything from the initial characterization of the materials to their impact on human health. However, assessing the toxicity of NMs remains challenging due to discrepancies in the available database of physicochemical and toxicological studies. An example that highlights the limited knowledge on the toxicity of NMs is the case of $TiO_2$ (E 171) in recent years. Although this substance has been approved as a food additive in the European Union (EU), the safety of its usage was re-evaluated by the Panel on Food Additives and Nutrient Sources added to Food (EFSA ANS) in 2016. The panel concluded that further research is needed to address gaps in knowledge regarding the potential effects of $TiO_2$ NPs on the reproductive system, in order to establish an Acceptable Daily Intake. Concerns were also raised about the characterization of the material used as a food additive, particularly regarding the particle size and distribution of $TiO_2$. The French Agency for Food, Environment, and Occupational Health Safety (ANSES) conducted a review of the risks associated with exposure to the food additive, which confirmed the uncertainties and data gaps previously identified by the EFSA. In the same year, the Netherlands Food and Consumer Product Safety Authority (NVWA) also provided an opinion on the potential health effects of $TiO_2$ as a food additive, emphasizing the need to consider both immunotoxicological and reprotoxicological effects.

This example illustrates how legislation has been reviewed for one of the most commonly used and well-known NMs, which is often described as a safe material. However, this becomes even more challenging when dealing with new nanoformulations, as each type of $MO_x$ NM has unique characteristics that require a thorough evaluation on a case-by-case basis. The limited information provided by current toxicology methodologies is a major problem and can pose significant difficulties for risk assessors. Traditional 2D static cultures and expensive animal studies are inadequate to provide meaningful insights into the toxicity of NMs on human tissues and organs. In vitro models using 2D static cell cultures do not adequately replicate the complex interactions between cells and their surrounding matrices, as well as the continuous fluid and blood circulation observed in native human organs. As a result, these models are inherently inaccurate for toxicity studies. Likewise, animal models fail to capture precise molecular mechanisms due to differences in physiological responses between animals and humans. However, recent advancements in tissue engineering and 3D cell culture offer promising alternatives for assessing NP toxicity, bridging the gap between preclinical models and human systems. One such alternative is the use of MicroPhysiological Systems (MPS), also known as Organ-On-a-Chip (OoC) platforms. These microfluidic systems cultivate human tissues in a 3D environment, providing biomimetic platforms that are more physiologically relevant compared to conventional models. Nowadays, various OoC models were developed, including those for the skin, intestine, liver, kidney, and lung, and their utility in assessing the toxicity of Metal and $MO_x$ NPs has already been considered (Lu and Radisic, 2021, Ashammakhi et al., 2019). For example, Zhang et al. (2018) developed a biomimetic 3D lung-on-a-chip to evaluate the pulmonary toxicity of ZnO and $TiO_2$ NPs. This study simultaneously analyzed cellular morphology, junction protein expression, ROS generation, and apoptosis of epithelial and endothelial cells exposed to NPs. More recently, a lung-on-a-chip platform using a physiologically relevant flow rate was employed to investigate the toxicity of ZnO (Arathi et al., 2022).

Moreover, in a study conducted by Ahn et al. in 2018, a Heart-On-Chip platform was developed using Mussel-inspired 3D fiber scaffolds to address the toxicity of engineered $TiO_2$ and Ag NPs. It was discovered that engineered NPs cause a reduction in the contractile function of cardiac tissues due to structural damage to the tissue architecture. Another example of assessing nanotoxicity in 3D cardiac tissue was described by Lu et al. in 2021. Using their 3D heart model, they observed that CuO NPs induce electrical and contractile dysfunction through the generation of ROS, while $SiO_2$ leads to the secretion of pro-inflammatory cytokines. A 3D Epidermal Model has also been developed to investigate the toxic mechanisms of silver NPs in studies conducted by Chen et al. in 2019 and Wills et al. in 2015. Additionally, Yin et al. in 2019 developed a 3D human placenta-on-a-chip model to monitor NP exposure at the placental barrier. Furthermore, Li et al. in 2019b created a 3D microfluidic hepatocyte platform for evaluating the hepatotoxicity of $Fe_3O_4$ NPs, revealing that cumulative exposure to magnetic NPs via the 3D hepatocyte chip results in significant damage to hepatocytes. The introduction of advanced OoC technology could improve the assessment of nanotoxicity for new nanoformulations by providing more relevant human data. Combining OoC technology with AI and in silico computational models, as suggested by Halder et al. in 2020, could potentially offer more predictive models for nanotoxicity assessment with a high correlation between in vitro and in vivo results, thus facilitating the development of new $MO_x$-based nanoformulations that have a positive impact in the agri-food sector.

## 5. Concluding remarks

The potential of $MO_x$ NPs for beneficiary changes in agriculture, food, and the environment has been highlighted through state-of-the-art research on their utilization in crop growth, water and soil remediation, and food quality control.

In agricultural production, MOx NMs in the formulation of nanopesticides and fertilizers have the potential to increase crop growth efficiency compared to conventional formulations while also providing solutions for wastewater treatment and soil remediation.

In the food sector, the possibility to efficiently incorporate $MO_x$ NMs within packaging offers benefits like increasing food shelf life, protecting products, and providing information on product quality through food safety sensors.

However, as with all new technologies, the potential benefits of $MO_x$ NMs must be balanced against risks, as concerns about their hazardous effects on human health and the environment persist. Humans may be exposed to $MO_x$ NPs in food through migration from packaging films and their accumulation in edible plant portions.

This is why commercial applications in real farmlands and food chains are still in the early stage. Therefore, a proper nanosafety assessment is urgently needed. Toxicological assessment of NPs on human health relies mainly on in vivo data from animal studies, which do not reflect human physiology and are difficult to implement due to the diversity in types of $MO_x$ NM. In parallel, there is still a lack of predictive in-vitro platforms as the standard 2D cell culture assays fail to mimic human physiology.

## Acknowledgment

The work described in this article has been conducted within the project MicroLabAptaSens. This project has received funding from the Science Fund of the Republic of Serbia, within the IDEAS programme under grant agreement No 7750276. This research was supported in part by the European Union's Horizon 2020 research and innovation programme under the Marie Skłodowska–Curie Grant Agreement No. 872662 (IPANEMA) and through ANTARES project that has received funding from the European Union's Horizon 2020 research and innovation programme under grant agreement SGA-CSA. No. 739570 under FPA No. 664387. This work was supported by the Science Fund of the Republic of Serbia (Grant No. 7747845, *In situ pollutants removal from waters by sustainable green nanotechnologies*-CleanNanoCatalyze) and by the Ministry of Education, Science and Technological Development of the Republic of Serbia (Grant No. 451-03-68/2022-14/200125).